\documentclass[12pt]{article}
\usepackage{amsfonts}
\usepackage{amsmath}
\newcommand{\be}{\begin{equation}}
\newcommand{\ee}{\end{equation}}
\newcommand{\bea}{\begin{eqnarray}}
\newcommand{\eea}{\end{eqnarray}}
\def \la{\label}

\def\({\left (}
\def\){\right )}
\def\]{\right]}
\def\<{\left <}
\def\>{\right>}

\newcommand{\bx}{\mathbf{x}}
\newcommand{\bk}{\mathbf{k}}
\newcommand{\by}{\mathbf{y}}
\newcommand{\A}{\mathbf{A}}
\newcommand{\bp}{\mathbf{p}}
\def \b\xi {{\pmb{\xi}}}
\newcommand{\bxi}{\pmb{\xi}}

\newcommand{\bs}{\pmb{\sigma}}
\newcommand{\bJ}{\pmb{\mathcal{J}}}
\newcommand{\bP}{\pmb{\mathcal{P}}}
\newcommand{\J}{\mathcal{J}}
\newcommand{\bj}{\mathbf{j}}
\newcommand{\F}{\mathcal{F}}
\newcommand{\D}{\mathrm{D}}
\newcommand{\bX}{\mathbf{X}}
\newcommand{\bR}{\mathbf{R}}
\newcommand{\cl}{\mathcal{L}}
\newcommand{\bB}{\mathbf{B}}

\newcommand{\Wc}{{W_\text{c}}}
\newcommand{\Wm}{{W_\text{m}}}

\newcommand{\Vc}{V_\text{c}}
\newcommand{\Vel}{{V_\text{elec}}}
\newcommand{\Phiel}{{\Phi_\text{elec}}}
\newcommand{\FR}{{F^\text{R}}}
\newcommand{\EL}{E_\text{l}}
\newcommand{\ET}{E_\text{t}}
\newcommand{\e}{\mathrm{e}}
\renewcommand{\r}{{\mathbf{r}}}
\renewcommand{\k}{{\mathbf{k}}}
\renewcommand{\d}{\mathrm{d}}
\newcommand{\Order}{\mathrm{O}}
\newcommand{\order}{\mathrm{o}}
\newcommand{\Tr}{\mathrm{Tr}}

\newcommand{\ie}{{\it i.e.}}

\begin{document}

\begin{center}

{\Large\bf Equilibrium correlations in charged fluids coupled to the radiation field}

\vspace{5mm}

Sami El Boustani, Pascal R. Buenzli,\footnote{Electronic address:
    Pascal.Buenzli@epfl.ch} {\normalsize and} Philippe
A. Martin\\ Institute of Theoretical Physics\\ Swiss Federal Institute of
Technology Lausanne\\ CH-1015, Lausanne EPFL, Switzerland\\

\end{center}

\vspace{5mm}

\abstract{We provide an exact microscopic statistical treatment of particle
    and field correlations in a system of quantum charges in equilibrium
    with a classical radiation field. Using the Feynman-Kac-It\^o
    representation of the Gibbs weight, the system of particles is mapped
    onto a collection of random charged wires. The field degrees of freedom
    can be integrated out, providing an effective pairwise magnetic
    potential. We then calculate the contribution of the transverse field
    coupling to the large-distance particle correlations. The asymptotics
    of the field correlations in the plasma are also exactly determined.}

\vspace{5mm}

\noindent PACS numbers~: \verb|05.30.-d|, \verb|05.40.-a|, \verb|11.10.Wx|

\section{Introduction}

Thermal states of non relativistic particles interacting by the sole
Coulomb potential are known to provide an adequate description of many
states of matter. The introduction of magnetic interactions between the
particles poses a novel problem since they are mediated by the coupling to
the transverse part of the electromagnetic field. This immediately leads to
consider the full system of matter in equilibrium with radiation : the
relevant theory becomes then the thermal quantum electrodynamics (thermal
QED).

In order to go beyond pure electrostatics without facing the full QED, a
number of studies rely on the Darwin approximation.  Darwin has shown
\cite{Darwin}, \cite{Landau} that one can eliminate the transverse degrees
of freedom of the field within the Lagrangian formalism up to order
$c^{-2}$ ($c$ is the speed of light). A nice review of the derivation of
the Darwin Lagrangian and a lucid discussion of its consequences can be
found in \cite{Essen}. The resulting Darwin Hamiltonian can be used to
investigate the equilibrium properties of the so called weakly relativistic
plasmas; see the recent works of Appel and Alastuey
\cite{Alastuey-Appel-Physica-238}, \cite{Appel-Alastuey-Physica-252},
\cite{Appel-Alastuey-Phys.Rev.}  and earlier references therein. These
authors have done a careful analysis of the domain of validity of the
Darwin approximation and shown in particular that the predictions of the
Darwin Hamiltonian on the tail of particle correlations in thermal states
cannot be correct. Indeed the well-known Bohr--van Leeuwen theorem
\cite{Alastuey-Appel-Physica-276} asserts that classical (non-quantum)
matter completely decouples from the radiation field. Thus the Darwin
Hamiltonian, which treats the particles classically, should not predict any
effect of the transverse field when used for thermal equilibrium
computations.  The Darwin approximation is, however, not deprived of any
meaning in statistical physics. Indeed, the authors show in
\cite{Appel-Alastuey-Phys.Rev.} that Darwin predictions about current
correlations coincide with those of thermal QED in the restricted window of
distances $\lambda_\text{part}\ll r \ll \lambda_\text{ph}$, where
$\lambda_\text{part}=\hbar\sqrt{\beta/m}$ is the de Broglie thermal
wavelength of the particles and $\lambda_\text{ph}=\beta\hbar c$ the
thermal wavelength of the photons.
But to determine the tail $r\gg \lambda_\text{ph}$ of the correlations in
the presence of the radiation field, matter has to be treated quantum
mechanically to avoid the conclusion of the Bohr--van Leeuwen theorem. The
situation is similar to orbital diamagnetism in equilibrium, which is of
quantum-mechanical origin.

In this work, we consider equilibrium states of non-relativistic spinless
quantum charges coupled with the radiation field in the standard way
(section 2).  We shall, however, treat the field classically on the ground
that the large distances $r\gg \lambda_\text{ph}$ are controlled by the
small wave numbers $k\sim \tfrac{1}{r}\ll\tfrac{1}{\lambda_\text{ph}}$,
implying $\beta\hbar \omega_{\bk}\sim \frac{\lambda_\text{ph}}{r}\ll
1$. Hence only long-wavelength photons will contribute to the asymptotics
which is expected to be adequately described by classical fields. The full
QED model with quantized electromagnetic field will be studied in a
subsequent work (see also comments in the concluding remarks, section 8).

Our main tool will be the Feynman-Kac-It\^o path integral representation of
the degrees of freedom of the charges.
The Feynman-Kac integral representation has been widely used to derive
various properties of quantum Coulomb systems, in particular to determine
the exact large-distance behaviour of the correlations; see \cite{Cornu1},
\cite{Cornu3}, and \cite{Alastuey}, \cite{Brydges-Martin} for reviews. In
this representation quantum charges become fluctuating charged loops
(closed Brownian paths), formally analogous to classical fluctuating wires
carrying multipoles of all orders. These fluctuations are responsible for
the lack of exponential screening in the quantum plasma and for an
algebraic tail $\sim r^{-6}$ of the particle correlations
\cite{Alastuey-Martin}.

Adding an external magnetic field produces a phase factor in the
Feynman-Kac-It\^o formula, whose argument is the flux of the magnetic field
across the random loop. Correlations in the case of an homogeneous external
magnetic field have been studied in \cite{Cornu2}.  When the particles are
thermalized with the field, the latter becomes itself random and
distributed according to the thermal weight of the free radiation. The
system can be viewed as a classical-like system of random loops immersed in
a random electromagnetic field.  At this point, the field degrees of
freedom can be exactly integrated out by means of a simple Gaussian
integral since the Hamiltonian of free radiation is quadratic in the field
amplitudes. One is then left with an effective pairwise current-current
interaction between the loops which has a form similar to the magnetostatic
energy between a pair of classical currents.  For the sake of illustrating
the basic mechanisms in a simple setting, this program is carried out in
section 3
with particles obeying Maxwell-Boltzmann statistics.
Appropriate modifications needed to take into account the particle
statistics (Bose or Fermi) are given in section 7.

In section 4 we apply the formalism to the determination of the asymptotic
form of the correlation between two quantum particles embedded in a
classical plasma. This simple model already illustrates the main features
occurring in the general system. The effective magnetic interaction
contributes to the $r^{-6}$ tail, but its ratio to the Coulombic
contribution is of the order of the square of the relativistic parameter
$(\beta mc^{2})^{-1}=(\lambda_\text{part}/\lambda_\text{ph})^2$.

In section 5 we consider the generalization of the results obtained for two
particles to the full system of quantum charges. The analysis relies on the
technique of quantum Mayer graphs previously developed for Coulomb systems,
and we merely point out the few changes that are needed to include the
effective magnetic interactions.

Field fluctuations in plasmas have been studied for a long time at
macroscopic scales, much larger than interparticle distances; see
\cite{Landau}, \cite{Felderhof} and references cited therein. In section 6,
we reexamine this question from a microscopic viewpoint and show that
electromagnetic field correlations are always long ranged due to the
quantum nature of the particles. This is in disagreement with the
prediction of macroscopic theories. We come back to this point in the
concluding remarks (section 8). However, in the classical limit, we recover
the fact already observed in \cite{Felderhof} that the long-range behaviour
of the longitudinal and transverse parts of the electric field correlations
compensate exactly.

In section 7, we generalise the formalism developed in section 3 to include
Bose and Fermi particle statistics. This is done as usual by decomposing
the permutation group into cycles and grouping particles belonging to a
cycle into an extended Brownian loop. When this is combined with the
Feynman-Kac-It\^o path integral representation of the particles, the system
takes again a classical-like form: a collection of Brownian loops immersed
in a classical random electromagnetic field.  At this point the physical
quantities can again be analyzed in terms of Mayer graphs comprising
pairwise Coulomb and effective magnetic interactions, as in section 5.

The methods presented in this paper have been applied to the study of the
semi-classical Casimir effect \cite{Buenzli-Martin},
\cite{Buenzli-Martin2}.

\section{The model}
We first consider the QED model for non-relativistic quantum charges
(electrons, nuclei, ions) with masses $m_{\gamma}$ and charges $e_{\gamma}$
contained in a box $\Lambda\in\mathbb{R}^3$ of linear size $L$ and
appropriate statistics. The index $\gamma$ labels the ${\cal S}$ different
species and runs from $1$ to ${\cal S}$.  The particles are in equilibrium
with the radiation field at temperature $T$.  The field is itself enclosed
into a large box $K$ with sides of length $R,\,R\gg L$.  The Hamiltonian of
the total finite volume system reads, in Gaussian units,
\begin{align}
    H_{L,R}=\sum_{i=1}^{n}\frac{\({\bf
            p}_{i}-\frac{e_{\gamma_{i}}}{c}\A(\r_{i})\)^{2}}{2m_{\gamma_{i}}}+
    \sum_{i<j}^{n}
    \frac{e_{\gamma_{i}}e_{\gamma_{j}}}{|\r_{i}-\r_{j}|}+\sum_{i=1}^{n} V_{{\rm
            walls}}(\gamma_{i},\r_{i})+H_{0}^{{\rm rad}}. \la{B.1}
\end{align}
The sums run on all particles with position $\r_{i}$, momentum
$\mathbf{p}_i$, and species index $\gamma_{i}$; $V_{{\rm
        walls}}(\gamma_{i},\r_{i})$ is a steep external potential that
confines a particle in $\Lambda$. It can eventually be taken infinitely
steep at the wall's position, implying Dirichlet boundary conditions---\ie,
vanishing of the particle wave functions at the boundaries of $\Lambda$.
The electromagnetic field is written in the Coulomb (or transverse) gauge
so that the vector potential $\A(\r)$ is divergence free and $H_{0}^{{\rm
        rad}}$ is the Hamiltonian of the free radiation field. The Coulomb
gauge is usually preferred for simplicity in situations where the particles
are non-relativistic and high-energy processes are neglected
\cite{Cohen}. It has the advantage to clearly disentangle electrostatic and
magnetic couplings in the Hamiltonian.

We impose periodic boundary conditions on the faces of the large box $K$
\footnote{Periodic conditions are convenient here. We could as well choose
    metallic boundary conditions. Since the field region $K$ will be
    extended over all space right away, the choice of conditions on the
    boundaries of $K$ are expected to make no differences for the particles
    confined in $\Lambda$.}.  Hence expanding $\A(\r)$ and the free photon
energy $H_{0}^{{\rm rad}}$ in the plane-wave modes $\bk= (\frac{2\pi
    n_{x}}{R},\frac{2\pi n_{y}}{R},\frac{2\pi n_{z}}{R})$ gives
\begin{align}
    \A(\r)&=\(\frac{4\pi \hbar c^{2}}{R^{3}}\)^{1/2}\sum_{\bk\lambda}g(\bk)
    \frac{ {\bf e}_{\bk\lambda}}{\sqrt{2\omega_{\bk}}}
    (a_{\bk\lambda}^{*}\e^{-i\bk\cdot\r}+a_{\bk\lambda}\e^{i\bk\cdot\r})\la{B.2}\\
    H_{0}^{{\rm rad}}&={\sum_{\bk\lambda}}\hbar
    \omega_\bk\,a_{\bk\lambda}^{*}a_{\bk\lambda} \la{B.2a}
\end{align}
where $a_{\bk\lambda}^{*}$ and $\,a_{\bk\lambda}$ are the creation and
annihilation operators for photons of modes $({\bk\lambda})$, ${\bf
    e}_{\bk\lambda}$ ($\lambda=1,2$) are two unit polarization vectors
orthogonal to ${\bk}$, and $\omega_\bk=ck,\; k=|\bk|$.  In (\ref{B.2}),
$g(\bk), \; g(0)=1$, is a real spherically symmetric smooth form factor
needed to take care of the ultraviolet divergencies. It is supposed to
decay rapidly beyond the characteristic wave number $k_c=m c/\hbar$ (see
\cite{Cohen}, chap. 3). Since we are interested in the large-distance
$r\to\infty$ asymptotics, related to the small-$\bk$ behaviour $k\to 0$,
the final result will be independent of this cut-off function.

The total partition function
\begin{align}
    Z_{L,R}= \Tr\ \e^{-\beta H_{L,R}}
    \la{B.3}
\end{align}
is obtained by carrying the trace $\Tr=\Tr_\text{mat} \Tr_\text{rad}$ of
the total Gibbs weight over particles' and the field's degrees of freedom~:
namely, on the particle wave functions with appropriate quantum statistics
and on the Fock states of the photons. The average values of observables
$\langle O_\text{mat}\rangle=Z_{L,R}^{-1}\Tr\( \e^{-\beta
    H_{L,R}}O_\text{mat}\)$ concerning only the particle degrees of freedom
can be computed from the reduced thermal weight
\begin{align}
    \rho_{L,R}=\frac{\Tr_\text{rad}\ \e^{-\beta H_{L,R}}}{Z^\text{rad}_{0, R}},
    \la{B.4}
\end{align}
where $Z_{0, R}^\text{rad}=\Tr_\text{rad}\exp{(-\beta H_{0}^{{\rm rad}})}$
is the partition function of the free radiation field, as follows from the
obvious identity
\begin{align}
    \langle O_\text{mat} \rangle= \frac{\Tr_\text{mat}
        \left(O_\text{mat}\,\rho_{L,R}\right)}{\Tr_\text{mat}\, \rho_{L,R}} \la{B.4a}.
\end{align}
We shall perform the thermodynamic limit in two stages by first letting
$R\to\infty$. Then $\rho_{L}=\lim_{R\to\infty}\rho_{L,R}$ defines the
effective statistical weight of the particles in $\Lambda$ immersed in an
infinitely extended thermalized radiation field.

As discussed in the Introduction, in this paper we treat the
electromagnetic field classically. This amounts to replacing the photon
creation and annihilation operators in (\ref{B.2}) and (\ref{B.2a}) by
complex amplitudes $\alpha_{\bk\lambda}^{*}$ and $\alpha_{\bk\lambda}$. In
this case, the free field distribution factorizes out as $\exp\({-\beta
    H_{R,L}}\)=\exp\({-\beta H_{0}^{{\rm rad}}}\)\exp\({-\beta
    H_\mathbf{A}}\)$, where
\begin{align}
    H_{R,L}=H_\mathbf{A}+H_{0}^{{\rm rad}},\quad H_\mathbf{A}= \sum_{i=1}^n \frac{\({\bf
            p}_{i}-\frac{e_{\gamma_{i}}}{c}\A(\r_{i})\)^{2}}{2m_{\gamma_{i}}}
    +U_\text{pot}(\r_{1},\gamma_{1},\ldots,\r_{n},\gamma_{n}), \la{B.4b}
\end{align}
and $U_\text{pot}$ is the total potential energy. Since the free radiation
weight $\exp\({-\beta H_{0}^{{\rm rad}}}\)$ is Gaussian,
$\A(\r)=\A(\r,\{\alpha_{\bk\lambda}\})$ can be viewed as a realization of a
Gaussian random field, and the term
$H_\mathbf{A}=H_{\A}(\{\alpha_{\bk\lambda}\})$ becomes the energy of the
particles in a given realization of the vector potential having Fourier
amplitudes $\{\alpha_{\bk\lambda}\}$.

The partial trace (\ref{B.4}) becomes, explicitly,
\begin{align}
    \rho_{L,R}=\left\langle  \e^{-\beta H_\mathbf{A}} \right\rangle_{{\rm rad}},
    \la{B.5}
\end{align}
where for a general function $F(\{\alpha_{\bk\lambda}\})$ of the mode
amplitudes $\left\langle F \right\rangle_{{\rm rad}}$ denotes the
normalized Gaussian average over all modes \footnote{The classical field is
    expanded as in (\ref{B.2}) and (\ref{B.2a}) with dimensionless
    amplitudes $\alpha_{\bk \lambda}$. In fact there will be no $\hbar$
    dependence arising from the field, as seen by changing everywhere
    $\alpha_{\bk\lambda} \mapsto \alpha_{\bk\lambda}/\sqrt{\hbar}$.}
\begin{align}
    \left\langle F \right\rangle_{{\rm rad}}=\prod_{\bk\lambda}\int
    \frac{\d^{2}\alpha_{\bk\lambda}}{\pi}\left[\beta\hbar \omega_{\bk}
        \e^{-\beta\hbar \omega_{\bk}|\alpha_{\bk \lambda}|^{2}}\right]
    F(\{\alpha_{\bk\lambda}\}). \la{B.5a}
\end{align}

Note that the stability of Coulombic matter and the existence of
thermodynamics for extended systems are assured if at least one of the
species obeys Fermi statistics \cite{Lieb}.  In the next section, merely as
a matter of simplifying the presentation, we compute the effective particle
interactions defined by $\rho_{L}$ ignoring quantum statistics.  In this
case, Maxwell-Boltzmann statistics requires the presence of an additional
short-range repulsive potential
$V_\text{sr}(\gamma_i,\gamma_j,|\r_i-\r_j|)$ in the Hamiltonian (\ref{B.1})
to prevent the collapse of opposite charges and guarantee thermodynamical
stability. The generalization to Fermi and Bose statistics will be given in
section 7.

\section{The gas of charged loops and the effective magnetic interaction}
We now introduce the Feynman-Kac-It\^o path integral representation of the
configurational matrix element $\langle \r_1,...,\r_n | \e^{-\beta H_{\A}}
| \r_1,..., \r_n\rangle $ for the particles interacting with a fixed
realization of the field. For a single particle of mass $m$ and charge $e$
in a scalar potential $V^\text{ext}(\r)$ and vector potential $\A(\r)$, we
first recall that this matrix element reads \cite{Feynman-Hibbs},
\cite{Roepstorff}, \cite{Simon}
\begin{align}
    &\langle \r |
    \exp\left(-\beta\left[\frac{\(\bp-\frac{e}{c}\A(\r)\)^{2}}{2m}+V^\text{ext}(\r)\right]\right)|\r\rangle=
    \(\frac{1}{2\pi\lambda^2}\)^{3/2} \!\int\!\! \D(\b\xi) \nonumber\\
    &\times
    \exp\left(-\beta\left[\int_{0}^{1}\!\!\! \d s\ V^\text{ext}\big(\r +\lambda \b\xi
        (s)\big)-i\frac{e}{\sqrt{\beta m c^{2}}}\int_{0}^{1} \!\!\! \d\b\xi(s) \cdot
        \A\big(\r+\lambda \b\xi (s)\big)\right] \right).
    \label{3.1}
\end{align}
Here $\b\xi (s),\;0\leq s \leq1,\; \b\xi (0)=\b\xi (1)=\mathbf{0}$, is a
closed dimensionless Brownian path and $\D(\b\xi)$ is the corresponding
conditional Wiener measure normalized to $1$. It is Gaussian, formally
written as
\\
$\exp\Big(-\frac{1}{2}\int_0^1 \d s \left|\frac{\d\b\xi (s)}{\d
    s}\right|^2\Big)\d[\b\xi (\cdot)]$, with zero mean and covariance
\begin{align}
    \int\!\! \D(\b\xi)\,\xi^{\mu}(s_1)\xi^{\nu}(s_2)=\delta^{\mu\nu}(\min(s_1,\:s_2)-s_1 s_2)
    \label{3.2}
\end{align}
where $\xi^{\mu}(s)$ are the Cartesian coordinates of $\b\xi(s)$.  In this
representation a quantum point charge looks like a classical charged closed
loop denoted by $\F=(\r,\b\xi)$, located at $\r$ and with a random shape
$\b\xi(s)$ having an extension given by the de Broglie length
$\lambda=\hbar\sqrt{\beta/m}$ (the quantum fluctuation).  The magnetic
phase in (\ref{3.1}) is a stochastic line integral: it is the flux of the
magnetic field across the closed loop. The correct interpretation of this
stochastic integral is given by the rule of the middle point; namely, the
integral on a small element of line $\bx-\bx'$ is defined by
\begin{align}
    \int_{\bx}^{\bx'}\!\!\!\d\b\xi\cdot {\bf f}(\b\xi)=(\bx-\bx')\cdot{\bf f}
    \(\frac{\bx+\bx'}{2}\),\quad \bx-\bx'\to 0 \la{3.3}
\end{align}
We shall stick to this rule when performing explicit
calculations.\footnote{Other prescriptions are possible for the path
    integral to correctly represent the quantum mechanical Gibbs weight in
    presence of a magnetic field. The It\^o rule may be used when ${\bf f}$
    is divergence free \cite{Roepstorff}.} Note the dimensionless
relativistic factor $(\beta m c^{2})^{-1/2}$ in front of the vector
potential term.

This is readily generalized to a system of $n$ interacting particles~: The
weight in the space of $n$ loops $\F_{1}=(\r_{1},\gamma_{1},
\b\xi_{1}),\ldots,\F_{n}=(\r_{n},\gamma_{n},\b\xi_{n})$ coming from the
path integral representation of \\ \mbox{$\langle \r_1,...,\r_n |
    \e^{-\beta H_{\A}} | \r_1,..., \r_n\rangle$} is $\exp(-\beta
U(\F_1,...,\F_n,\A))$ where
\begin{align}
    U(\F_1,...,\F_n, \A) =&\sum_{i<j}^{n}e_{\gamma_{i}}e_{\gamma_{j}}\Vc
        (\F_{i},\F_{j}) \nonumber
        \\&-i\sum_{j=1}^{n}\frac{e_{\gamma_{j}}} {\sqrt{\beta
                m_{\gamma_{j}}c^{2}}} \int_{0}^{1}\!\!\! \d\b\xi_j(s) \cdot
        \A(\r_j+\lambda_{\gamma_j} \b\xi_j (s)) \la{3.4}
\end{align}
The matrix element $\langle \r_1,...,\r_n | \e^{-\beta H_{\A}} | \r_1,...,
\r_n\rangle $ is obtained by integrating $\exp(-\beta U(\F_1,...,\F_n,\A))$
over the random shapes $\bxi_1,...,\bxi_n$ of the loops, as in (\ref{3.1}).
In (\ref{3.4}),
\begin{align}
    \Vc(\F_{i},\F_{j})=\int_{0}^{1}\!\!\d s\ \frac{1}
    {|\r_{i}+\lambda_{\gamma_{i}}\b\xi _{i}(s)-\r_{j}-\lambda_{\gamma_{j}}\b\xi
        _{j}(s)|} \la{B.6}
\end{align}
is the Coulomb potential between two loops, and for the sake of brevity, we
have omitted the non electromagnetic terms
\begin{align}
	\sum_{i<j}^n V_\text{sr}(\F_i,\F_j) + \sum_{i=1}^n V_\text{walls}(\F_i) \label{non-em-terms}
\end{align}
corresponding to the short-range regularization and to the confinement
potential. The vector potential term can be written as \\ \mbox{$-i\int\!\!
    \d\bx\ \A(\bx)\cdot \bJ(\bx)$} in terms of current densities associated
with the Brownian loops~:
\begin{align}
    \bJ(\bx)=\sum_{i=1}^{n}\bj(\F_i,\bx),\quad
    \bj(\F_i,\bx)=\frac{e_{\gamma_{i}}}{\sqrt{\beta m_{\gamma_i} c^{2}}}\int_{0}^{1}
    \!\!\!\d\b\xi_{i}(s)\ \delta(\bx-\r_{i}-\lambda_{\gamma_{i}}\b\xi_{i}(s)).
    \la{B.8}
\end{align}
If one interprets the (ill-defined) derivative
$\lambda_{\gamma_{i}}\d\b\xi_{i}(s)/\d s={\bf v}_{i}(s)$ as the
``velocity'' of a particle of charge $e_{\gamma_{i}}$ moving along the loop
$\b\xi_{i}(s)$, the quantity $e_{\gamma_{i}}{\bf
    v}_{i}(s)\delta(\bx-\r_{i}-\lambda_{\gamma_{i}}\b\xi_{i}(s)) $
corresponds to a classical current density. This is just a formal
analogy. In subsequent calculations of stochastic integrals arising from
(\ref{B.8}), we will always use the mathematically well-defined rule of the
middle point (\ref{3.3}).  Moreover, such ``imaginary time'' currents
appearing in the Feynman-Kac-It\^o representation are not the physical
``real-time'' current observables. Our definition (\ref{B.8}) also includes
the relativistic factor $(\beta m_{\gamma_i} c^{2})^{-1/2}$.

A remarkable fact is that the transverse part of the field enters in
\mbox{$\exp(-\beta U(\F_1,...,\F_n,\A))$} as a phase factor linear in
$\A$ and its Fourier amplitudes (contrary to the Hamiltonian (\ref{B.1})
written in operatorial form). Since the statistical weight $\e^{-\beta
    H_{0}^{{\rm rad}}}$ (\ref{B.2a}) is a Gaussian function of these
Fourier amplitudes, it makes it possible to perform explicitly the partial
trace over the field degrees of freedom in (\ref{B.5}) according to the
following steps~:
\begin{align}
    &\left\langle\exp\left[i\beta\int\!\! \d\bx\ \A(\bx)\cdot \bJ(\bx)\right]
    \right\rangle_{{\rm rad}}= \left\langle\prod_{\bk\lambda}\exp
    \left[i(u_{\bk\lambda}^{*}\alpha_{\bk\lambda}+u_{\bk\lambda}\alpha_{\bk\lambda}^{*})\right]
    \right\rangle_{{\rm rad}}=\nonumber\\
    &\exp\left[-\frac{\beta}{2R^{3}}\sum_{\bk\lambda}\frac{4\pi
            g^{2}(\bk)}{k^{2}}\left|\bJ(\bk)\cdot{\bf
            e}_{\bk\lambda}\right|^{2}\right] =
    \exp\left[-\frac{\beta}{2}\int\!\!\frac{\d\bk}{(2\pi)^{3}}
        (\J^\mu(\bk))^{*}G^{\mu\nu}(\bk)\J^\nu(\bk)\right].
    \la{B.9}
\end{align}
The first equality is obtained by introducing the mode expansion
(\ref{B.2}), yielding
\begin{align}
    u_{\bk\lambda}=\beta\(\frac{4\pi\hbar c^{2}}{R^{3}}\)^{1/2}
    \frac{g(\bk)}{\sqrt{2\omega_{\bk}}}\ \bJ(\bk)\cdot{\bf e}_{\bk\lambda}, \quad
    \bJ(\bk)=\int\!\! \d\bx\ \e^{-i\bk\cdot\bx}\bJ(\bx) \la{B.10}
\end{align}
The second equality results from (\ref{B.5}), (\ref{B.5a}) and the Gaussian
integral
\mbox{$\int\!\frac{d^{2}\alpha}{\pi}\,\e^{-b|\alpha|^{2}+i(u^{*}\alpha+u\alpha^{*})}
    = b^{-1}\e^{-b^{-1}|u|^{2}}$}, $b>0$, whereas the infinite volume limit
$R\to\infty$ and the polarization sum have been performed in the last
equality. We have denoted by $G^{\mu\nu}(\k)$ the covariance of the free
transverse field~:
\begin{align}
    G^{\mu\nu}(\bk)=\frac{4\pi g^{2}(\bk)}{k^{2}}\delta_{{\rm
            tr}}^{\mu\nu}(\bk), \quad \delta_{{\rm
            tr}}^{\mu\nu}(\bk)=\delta^{\mu\nu}-\frac{k^{\mu}k^{\nu}}{k^{2}}, \quad k^\mu G^{\mu\nu}(\k) \equiv 0
    \la{B.11}
\end{align}
($\delta_{{\rm tr}}^{\mu\nu}(\bk)$ is the transverse Kronecker symbol).  In
(\ref{B.9}) and throughout the paper, summation on repeated vector
components $\mu,\nu=1,2,3$ is understood. In the configuration space, the
asymptotic behaviour of $G^{\mu\nu}(\bx)$ is obtained by approximating
$g^{2}(\bk)\sim 1$ in the inverse Fourier transform of $G^{\mu\nu}(\bk)$:
\begin{align}
    G^{\mu\nu}(\bx)\sim\int\!\! \frac{\d\bk
    }{(2\pi)^{3}}\,\e^{i\bk\cdot\bx}\,\frac{4\pi}{k^{2}}
    \(\delta^{\mu\nu}-\frac{k^{\mu}k^{\nu}}{k^{2}}\)
    =\frac{1}{2r}\(\delta_{\mu\nu}+\frac{x^\mu
        x^{\nu}}{r^{2}}\),\;r=|\bx|\to\infty \la{B.12}.
\end{align}
Decomposing the total current (\ref{B.8}) into the individual loop currents
we see that the effective weight (\ref{B.9}) takes the form
\begin{align}
    \left\langle\exp\left[i\beta\int\!\! \d\bx\ \A(\bx)\cdot \bJ(\bx)\right]
    \right\rangle_{{\rm rad}}=&\prod_{i=1}^{n}\exp\(-\frac{\beta
        e_{\gamma_{i}}^{2}}{2}\Wm(i,i)\)\nonumber\\
    &\times\exp\(-\beta\sum_{i<j}^{n}e_{\gamma_{i}}e_{\gamma_{j}}\Wm(i,j)\),
    \la{B.13}
\end{align}
where for two loops $i=\F_{i}$ and $j=\F_{j}$ we have introduced the
loop-loop effective magnetic potential
\begin{align}
    &e_{\gamma_{i}}e_{\gamma_{j}}\Wm(i,j)=\int\!\! \d\bx \!\!\int
    \!\!\d\by\ (j^{\mu}(\F_i,\bx))^{*}G^{\mu\nu}(\bx-\by)j^{\nu}(\F_j,\by)=
    \la{B.14}\\
    &=\frac{e_{\gamma_{i}}e_{\gamma_{j}}}{\beta
        \sqrt{m_{\gamma_i}m_{\gamma_j}}c^{2}}\int\!\! \frac{\d\bk }{(2\pi)^{3}}\,
    \e^{i\bk\cdot(\r_i-\r_j)}\!\!\int_{0}^{1}\!\!\d\xi_{i}^\mu(s_1)\,
    \e^{i\bk\cdot\lambda_{\gamma_{i}}\b\xi_{i}(s_1)}\!\! \int_{0}^{1}\!\!
    \d\xi_{j}^\nu(s_2)\, \e^{-i\bk\cdot\lambda_{\gamma_{j}}\b\xi_{j}(s_2)}
    G^{\mu\nu}(\bk). \nonumber
\end{align}
As a consequence of Gaussian integration, one recovers pairwise
interactions (\ref{B.14}) between loops.  The product in (\ref{B.13})
contains the magnetic self-energies of the loops.

It is pleasing and convenient that after averaging over the field modes,
the energy of the system of loops becomes an exact and explicit sum of pair
potentials (and self-energies)\footnote{We omit again in (\ref{loop-Gibbs})
    the non-electromagnetic terms (\ref{non-em-terms}).}~:
\begin{align}
	\left\langle \e^{-\beta U(\F_1,...,\F_n,\A)} \right\rangle_\text{rad} =
    \Big[\prod_{i=1}^{n}\e^{-\frac{\beta
                e_{\gamma_{i}}^{2}}{2}\Wm(i,i)}\Big]\ \e^{-\beta \sum_{i<j}
        e_{\gamma_i}e_{\gamma_j}\big( \Vc(i,j) + \Wm(i,j)
        \big)}. \label{loop-Gibbs}
\end{align}

It is interesting to ask for the status of the partial density matrix
(\ref{B.4}) compared to that generated by the Darwin Hamiltonian
$\rho_\text{Darwin} \propto \e^{-\beta H_\text{Darwin}}$ or, more
generally, if $\rho_{L,R}$ can be cast in the form $\rho_{L,R}\propto
\e^{-\beta H_\text{eff}}$ for some tractable Hamiltonian
$H_\text{eff}(\{\bp_{i},\r_{i}\})$ depending on the canonical variables of
the particles. The answer to this last question is very presumably
negative. Indeed the magnetic interaction (\ref{B.14}) is a two times
functional of the Brownian loops; namely, it lacks the equal-time constraint
occurring in the Coulomb potential (\ref{B.6}) (see the discussion before
(\ref{B.17}) below) necessary to come back to a simple operator form by
using the Feynman-Kac-It\^o formula backwards. This is a well-known common
feature of interactions resulting from integrating out external degrees of
freedom \cite{Feynman-Hibbs}.

The long-distance asymptotics of $\Wm(i,j)$ as $|\r_{i}-\r_{j}|\to\infty$ is
determined by the small $\bk$ behaviour in the integrand of (\ref{B.14}).
Noting that $\int_{0}^{1}\d\b\xi(s)=0$ for a closed loop (It\^o's lemma), one
has
\begin{align}
    \int_{0}^{1}\!\!\d\xi_{i}^\mu(s)\,
    \e^{i\bk\cdot\lambda_{\gamma_{i}}\b\xi_{i}(s)} \sim i
    \lambda_{\gamma_i}
    \int_{0}^{1}\!\!\d\xi_{i}^\mu(s)\ \bk\cdot\b\xi_{i}(s),\quad \bk\to 0,
\end{align}
and thus
\begin{align}
    &\Wm(i,j)\sim
    \la{B.16}\\
    &\sim\frac{\lambda_{\gamma_{i}} \lambda_{\gamma_{j}}}{\beta
        \sqrt{m_{\gamma_{i}} m_{\gamma_{j}}}c^{2}} \!\int\!\!\! \frac{\d\bk
    }{(2\pi)^{3}}\e^{i\bk\cdot(\r_{i}-\r_{j})} \!\!\int_{0}^{1}\!\!\! \d\xi^\mu_{i}(s_{1})
    (\bk\cdot\b\xi_{i}(s_{1})) \!\!\int_{0}^{1}\!\!\! \d\xi^\nu_{j}(s_{2})
    (\bk\cdot\b\xi_{j}(s_{2}))G^{\mu\nu}(\bk)\nonumber\\
    &=\frac{\lambda_{\gamma_{i}} \lambda_{\gamma_{j}}}{\beta
        \sqrt{m_{\gamma_{i}} m_{\gamma_{j}}}c^{2}} \!\int_{0}^{1}\!\!\!
    \d\xi_{i}^{\mu}(s_{1}) (\b\xi_{i}(s_{1})\cdot {\bf \nabla}_{\r_{i}})
    \!\!\int_{0}^{1}\!\!\! \d\xi_{j}^{\nu}(s_{2}) (\b\xi_{j}(s_{2})\cdot{\bf
        \nabla}_{\r_{j}})\,G^{\mu\nu}(\r_{i}-\r_{j}),\nonumber
\end{align}
as $ |\r_{i}-\r_{j}|\to \infty$.  Upon using the asymptotic form
(\ref{B.12}) of $G^{\mu\nu}(\r_{i}-\r_{j})$, it is clear that for fixed
loop shapes $\b\xi_{i}$ and $\b\xi_{j}$ the decay of $\Wm(i,j)$ is $\sim
|\r_{i}-\r_{j}|^{-3}$.  It is of dipolar type modified by the constraint
imposed by the transversality.

The Coulombic part (\ref{B.6}) of the loop-loop interaction still decays as
$r^{-1}$ and deserves the following remark. From the Feynman-Kac formula
the potential (\ref{B.6}) inherits the quantum-mechanical equal-time
constraint; \ie, every element of charge
$e_{\gamma_{i}}\lambda_{\gamma_{i}}\d\b\xi_i(s_{1})$ of the first loop does
not interact with every other element
$e_{\gamma_{j}}\lambda_{\gamma_{j}}\d\b\xi_j(s_{2})$ as would be the case
in classical physics, but the interaction takes place only if $s_1
=s_2$. It is therefore of interest to split
\begin{align}
    \Vc(i,j)=\Vel(i,j)+\Wc(i,j), \la{B.17}
\end{align}
where
\begin{align}
    \Vel(i,j)= \int_{0}^{1}\!\!\! \d s_{1} \!\int_{0}^{1}\!\!\! \d
    s_{2}\ \frac{1} {|\r_{i}+\lambda_{\gamma_{i}}\b\xi
        _{i}(s_{1})-\r_{j}-\lambda_{\gamma_{j}}\b\xi _{j}(s_{2})|} \la{B.18}
\end{align}
is a genuine classical electrostatic potential between two charged loops
and
\begin{align}
    \Wc(i,j)=\int_{0}^{1}\!\!\! \d s_{1} \!\int_{0}^{1}\!\!\! \d
    s_{2}\,(\delta(s_{1}\!-\!s_{2})\!-\!1)\frac{1}
    {|\r_{i}+\lambda_{\gamma_{i}}\b\xi
        _{i}(s_{1})-\r_{j}-\lambda_{\gamma_{j}}\b\xi _{j}(s_{2})|} \la{B.19}
\end{align}
is the part of $\Vc(i,j)$ due to intrinsic quantum fluctuations ($\Wc(i,j)$
vanishes if $\hbar$ is set equal to zero). Because of the identities
\begin{align}
    \int_{0}^{1}\!\!\!\d s_{1}\,(\delta(s_{1}-s_{2})-1) = \int_{0}^{1}\!\!\! \d
    s_{2}\,(\delta(s_{1}-s_{2})-1)=0, \la{B.20}
\end{align}
the large-distance behaviour of $\Wc$ originates again from the term
bilinear in $\b\xi_{i}$ and $\b\xi_{j}$ in the multipolar expansion of the
Coulomb potential in (\ref{B.19})
\begin{align}
    &\Wc(i,j)\sim \la{B.21}
     \int_{0}^{1}\!\!\! \d s_{1}\!\! \int_{0}^{1}\!\!\! \d s_{2}\,
    (\delta(s_{1}\!-\!s_{2})\!-\!1)\(\lambda_{\gamma_{i}}\b\xi_{i}(s_{1})\cdot
    \nabla_{\r_{i}} \)\(
    \lambda_{\gamma_{j}}\b\xi_{j}(s_{2})\cdot\nabla_{\r_{j}}\)
    \frac{1}{|\r_{i}-\r_{j}|}.
\end{align}
It is dipolar and formally similar to that of two electrical dipoles of
sizes $e_{\gamma_{i}}\lambda_{\gamma_{i}}\b\xi_{i}$ and
$e_{\gamma_{j}}\lambda_{\gamma_{j}}\b\xi_{j}$.


\section{Two quantum charges in a classical plasma}

In order to exhibit the effect of the magnetic potential on the particle
correlations, we consider the simple model of two quantum charges $e_a$ and
$e_b$ with corresponding loops $\F_a=(\r_{a},\b\xi_{a})$ and
$\F_b=(\r_{b},\b\xi_{b})$ immersed in a configuration $\omega$ of classical
charges, following section VII of \cite{Alastuey-Martin} or section IV.C of
\cite{Brydges-Martin}.  According to (\ref{B.17}) one can decompose the
total energy as $\mathcal{U}(\F_{a},\F_{b},\omega)=e_a e_b
W(\F_{a},\F_{b})+U_\text{cl}(\F_{a},\F_{b},\omega) $ where
$W(\F_a,\F_b)=\Wc(\F_a,\F_b)+\Wm(\F_a,\F_b)$ is the sum of the electric and
magnetic quantum dipolar interactions and $U_\text{cl}(\F_a,\F_b,\omega)$
is the purely classical Coulomb energy (\ref{B.18}) of the two loops $\F_a$
and $\F_b$ together with that of the particles in the configuration
$\omega$. The correlation $\rho(\F_a,\F_b)$ between the loops is obtained
by integrating out the coordinates $\omega$ of the classical charges~:
\begin{align}
    \rho(\mathcal{F}_{a},\mathcal{F}_{b})=\frac{1}{\Xi_\text{cl}}\int_{\Lambda}\!\!\!
    \d\omega\
    \e^{-\beta\mathcal{U}(\mathcal{F}_{a},\mathcal{F}_{b},\omega)}=\e^{-\beta
        e_a e_b
        W(\mathcal{F}_{a},\mathcal{F}_{b})}\rho_\text{cl}(\mathcal{F}_{a},\mathcal{F}_{b}),
\label{B.22}
\end{align}
where $\Xi_\text{cl}$ is the partition function of the classical plasma and
$\rho_\text{cl}(\mathcal{F}_{a},\mathcal{F}_{b})$ is the correlation of the
two loops embedded in the plasma interacting with genuine classical Coulomb
forces. In the latter quantity, the classical theory of screening applies
so that effective interaction between the loops decay exponentially fast
\footnote{The usual Debye theory of screening has been rigorously shown to be
    valid at least at sufficiently high temperature
    \cite{Brydges-Federbush}.}. Thus one can approximate
$\rho_\text{cl}(\mathcal{F}_{a},\mathcal{F}_{b})$ in (\ref{B.22}) by
$\rho(\mathcal{F}_{a})\rho(\mathcal{F}_{b})$ up to a term exponentially decaying
as $|\r_{a}-\r_{b}|\to\infty$. Furthermore, integrating
$\rho(\mathcal{F}_{a},\mathcal{F}_{b})$ on the loop shapes leads to the
following expression for the positional correlation of the quantum charges
\begin{align}
& \rho(\r_{a},\r_{b})=
    \int\!\! \D(\b\xi_{a}) \!\!\int\!\! \D(\b\xi_{b})\ \e^{-\beta e_a e_b
        W(\mathcal{F}_{a},\mathcal{F}_{b})}\rho(\mathcal{F}_{a})
    \rho(\mathcal{F}_{b}) + \mathcal{O}(\e^{-C|\r_{a}-\r_{b}|})= \nonumber \\
 & = \rho_{a}\rho_{b}-\beta e_a e_b
    \int\!\! \D(\b\xi_{a}) \!\!\int\!\! \D(\b\xi_{b})\ 
    W(\mathcal{F}_{a},\mathcal{F}_{b}) \rho(\b\xi_{a})\rho(\b\xi_{b}) +
    \nonumber \\
 & \phantom{=} + \frac{1}{2}\beta^{2}e_a^2 e_b^2 \int\!\! \D(\b\xi_{a})
    \!\!\int\!\! \D(\b\xi_{b})\ W^{2}(\mathcal{F}_{a},\mathcal{F}_{b})
    \rho(\b\xi_{a})\rho(\b\xi_{b}) +...+
    \mathcal{O}(\e^{-C|\r_{a}-\r_{b}|})
    \la{B.23}
\end{align}
Since $W(\mathcal{F}_{a},\mathcal{F}_{b})\sim |\r_{a}-\r_{b}|^{-3}$ (see
(\ref{B.16}), (\ref{B.21})), the above expansion in powers of $W$ generates
algebraically decaying terms at large separation.  It is known that in a
homogeneous and isotropic phase, the electric dipole part $\Wc$ does not
contribute at linear order \cite{Alastuey-Martin},
\cite{Brydges-Martin}. The same is true for the magnetic part.  To see
this, it is convenient to write the linear $\Wm$ term of (\ref{B.23}) as
\begin{align}
    &-\beta e_a e_b \int\!\! \D(\b\xi_{a}) \!\!\int\!\! \D(\b\xi_{b})\ 
    \Wm(\mathcal{F}_{a},\mathcal{F}_{b}) \rho(\b\xi_{a}) \rho(\b\xi_{b})
    \nonumber\\
    &=-\frac{\beta e_a e_b}{\sqrt{\beta m_{a}c^{2}} \sqrt{\beta m_{b}c^{2}}}
    \int\!\!\! \frac{d^{3}\bk}{(2\pi)^{3}}\,
    \e^{i\bk\cdot(\r_{a}-\r_{b})}\ {t_{a}^{\mu}}^\ast(\bk) t_{b}^{\nu}(\bk)
    G^{\mu\nu}(\bk) \la{B.24}.
\end{align}
The stochastic $\b\xi_{a}$-line-integral is now included in the definition
of the tensor
\begin{align}
    t_{a}^{\mu}(\bk)= \int\!\! \D(\b\xi_{a})\rho(\b\xi_{a})
    \!\int_{0}^{1}\!\!\!\d\xi_{a}^{\mu}(s)\ \e^{-i\lambda_{a}\bk\cdot \b\xi_{a}(s)}
    \la{B.25}
\end{align}
and likewise for $t_{b}^{\nu}(\bk)$. Since both the measure $\D(\b\xi_{a})$
and $\rho(\b\xi_{a})$ are invariant under a rotation of $\b\xi_{a}$ in an
isotropic system, $t_{a}^{\mu}(\bk)$ transforms in a covariant manner under
rotations of $\bk$. Thus it is necessarily of the form
$t_{a}^{\mu}(\bk)=k^{\mu}f_{a}(|\bk|)$, implying the vanishing of
(\ref{B.24}) because of the transversality of $G^{\mu\nu}(\bk)$.  One
concludes that the slowest non-vanishing contribution comes from the
$W^{2}$ term in (\ref{B.23})
\begin{align}
 \rho(\r_{a},\r_{b})- \rho_{a} \rho_{b}=
 \frac{A(\beta)}{|\r_{a}-\r_{b}|^{6}}+\Order\(\frac{1}{|\r_{a}-\r_{b}|^{8}}\)  \la{B.26aaa}.
\end{align}
The temperature-dependent amplitude
$A(\beta)=A_\text{cc}(\beta)+A_\text{mm}(\beta)+A_\text{cm}(\beta)$
involves in principle electric and magnetic contributions from $\Wc^{2}$
and $\Wm^{2}$, as well as a cross contribution from $2\Wc\Wm$.  These
contributions can be calculated explicitly at lowest order in $\hbar$ (or
equivalently in the high-temperature limit $\beta\to 0$).  The electric
contribution in this limit is known to be \cite{Alastuey-Martin},
\cite{Brydges-Martin}
\begin{align}
    A_\text{cc}(\beta)\sim
    \hbar^{4}\frac{\beta^{4}}{240}\frac{e_{a}^{2}e_{b}^{2}}{m_{a}m_{b}}\rho_{a}\rho_{b}.
    \la{B.27}
\end{align}
To compute the magnetic contribution in the same limit, we write
the quadratic term
\begin{align}
    & \frac{\beta^{2} e_a^2 e_b^2}{2}\int\!\! \D(\b\xi_{a})\rho(\b\xi_{a})\!\!
    \int\!\!\D(\b\xi_{b})
    \rho(\b\xi_{b})\ \Wm^{2}(\mathcal{F}_{a},\mathcal{F}_{b})
    =\frac{e_{a}^{2}e_{b}^{2}}{2 m_{a}c^{2}m_{b}c^{2}}\nonumber\\
    & \times \int\!\!\! \frac{d^{3}\bk_{1}}{(2\pi)^{3}} \!\int\!\!\!
    \frac{d^{3}\bk_{2}}{(2\pi)^{3}}\, \e^{i(\bk_{1}+\bk_{2})\cdot(\r_{a}-\r_{b})}
    \,\left(T_{a}^{\mu\nu}(\bk_{1},\bk_{2})\right)^\ast {T_{b}^{\sigma\tau}}(\bk_{1},\bk_{2})
    G^{\mu\sigma}(\bk_{1}) G^{\nu\tau}(\bk_{2})
    \la{B.28}
 \end{align}
in terms of the tensors
\begin{align}
    T_{a}^{\mu\nu}(\bk_{1},\bk_{2})=\int\!\!\D(\b\xi_{a}) \rho(\b\xi_{a})
    \!\!\int_{0}^{1}\!\!\! \d\xi_{a}^{\mu}(s_{1}) \!\!\int_{0}^{1}\!\!\!
    \d\xi_{a}^{\nu}(s_{2})\ \e^{-i\lambda_{a}\bk_{1}\cdot\b\xi_{a}(s_{1})}
    \e^{-i\lambda_{a}\bk_{2} \cdot \b\xi_{a}(s_{2})} \la{B.29}
\end{align}
and $T_{b}^{\sigma\tau}(\bk_{1},\bk_{2})$, defined likewise.  As usual the
behaviour at large distances is controlled by that of the integrand of
(\ref{B.28}) at small wave numbers. Expanding (\ref{B.29}) at lowest order
in $\bk_{1}$ and $\bk_{2}$ gives
\begin{align}
    T_{a}^{\mu\nu}(\bk_{1},\bk_{2}) & \sim \int\!\! \D(\b\xi_{a})
    \rho(\b\xi_{a}) \!\!\int_{0}^{1}\!\!\! \d\xi_{a}^{\mu}(s_{1})
    \!\!\int_{0}^{1}\!\!\! \d\xi_{a}^{\nu}(s_{2})
    \left(-i\lambda_{a}\bk_{1}\cdot\b\xi_{a}(s_{1})\right)
    \left(-i\lambda_{a}\bk_{2}\cdot\b\xi_{a}(s_{2})\right) \nonumber \\
    & = -\lambda_{a}^{2}k_{1}^{\epsilon}k_{2}^{\eta}
    \int\!\! \D(\b\xi_{a})\rho(\b\xi_{a}) \!\!\int_{0}^{1}\!\!\! \d\xi_{a}^{\mu}(s_{1})
    \!\!\int_{0}^{1}\!\!\! \d\xi_{a}^{\nu}(s_{2})\ \xi_{a}^{\epsilon}(s_{1})
    \xi_{a}^{\eta}(s_{2})
\label{B.30}
\end{align}
and likewise for $T_{b}^{\sigma\tau}(\bk_{1},\bk_{2})$.  One sees that
because of the factor $\lambda_{a}^{2}\lambda_{b}^{2}$, the overall
contribution in (\ref{B.28}) will have a $\hbar^{4}$ factor so that at this
order we can neglect the quantum fluctuation in the density setting
$\rho(\b\xi_{a})\sim \rho_{a}$ independent of $\b\xi_{a}$. Thus the
stochastic integral to be calculated becomes (appendix A)
\begin{align}
    \int\!\!\D(\b\xi) \!\!\int_{0}^{1}\!\!\! \d\xi^{\mu}(s)
    \!\!\int_{0}^{1}\!\!\! \d\xi^{\nu}(t)\ \xi^{\epsilon}(s) \xi^{\eta}(t) =
    \frac{1}{12}(\delta^{\mu\nu}\delta^{\eta\epsilon} -
    \delta^{\mu\eta}\delta^{\nu\epsilon}),
\label{B.31}
\end{align}
leading to
\begin{align}
    T_{a}^{\mu\nu}(\bk_{1},\bk_{2}) &\sim -\frac{\lambda_{a}^{2}
        \rho_{a}}{12}(\delta^{\mu\nu}\bk_{1} \cdot \bk_{2}-k_{2}^{\mu}
    k_{1}^{\nu}),\nonumber\\
    T_{b}^{\sigma\tau}(\bk_{1},\bk_{2}) &\sim
    -\frac{\lambda_{b}^{2}\rho_{b}}{12}(\delta^{\sigma\tau}\bk_{1} \cdot
    \bk_{2}-k_{2}^{\sigma} k_{1}^{\tau}).
\la{B.32}
\end{align}
When this is inserted into (\ref{B.28}) and summation on vectorial indices
are performed, one finds the expression
\begin{align}
    A\int\!\!\! \frac{\d\bk_{1}}{(2\pi)^{3}} \!\!\int\!\!\!
    \frac{\d\bk_{2}}{(2\pi)^{3}}\, \e^{i(\bk_{1}+\bk_{2})\cdot(\r_{a}-\r_{b})}
    \,(4\pi)^{2} |g(\bk_{1})|^{2} |g(\bk_{2})|^{2}
    \left[1+\frac{(\bk_{1}\cdot\bk_{2})^{2}}{k_{1}^{2}k_{2}^{2}} \right],
    \la{B.33}
\end{align}
with $A=\frac{\lambda_{a}^{2} \lambda_{b}^{2} e_{a}^{2} e_{b}^{2} \rho_{a}
    \rho_{b}}{288m_{a}m_{b}c^{4}}$. The first term in the large brackets
gives a rapidly decaying contribution since it involves the Fourier
transform of the form factor $g^{2}(\bk)$. The algebraic large-distance
contribution comes from the second term which reads, after Fourier
transformation (approximating $g(\bk)\sim 1$, $\bk\to 0$),
\begin{align}
    A\(\partial_{\mu}\partial_{\nu}\frac{1}{|\r_{a}-\r_{b}|}\)
    \(\partial_{\mu}\partial_{\nu}\frac{1}{|\r_{a}-\r_{b}|}\)
    =A\;\frac{6}{|\r_{a}-\r_{b}|^{6}}. \la{B.34}
\end{align}
Finally one checks that there is no cross Coulomb-magnetic contribution
$A_\text{cm}(\beta)$ at the dominant order $\r^{-6}$ as a consequence of
transversality (appendix B). So adding (\ref{B.27}) and (\ref{B.34}) gives
the final result
\begin{align}
    \rho(\r_{a},\r_{b})- \rho_{a}\rho_{b} \sim \hbar^{4}\beta^{4}\frac{\rho_{a}
        \rho_{b}e_{a}^{2}e_{b}^{2}}{240 \,m_{a}m_{b}}\left[ 1+\frac{5}{(\beta
            m_{a}c^{2})(\beta m_{b}c^{2})}\right]\frac{1}{|\r_{a}-\r_{b}|^{6}}
    \la{B.26}
\end{align}
as $|\r_{a}-\r_{b}|\to\infty$ and at lowest order in $\hbar$.  One sees from
(\ref{B.6}) and (\ref{B.14}) that the order of magnitude of the ratio $\Wm/\Vc$ is
$(\beta m c^2)^{-1}$. In an electrolyte at room temperature $T=300 K$, this
ratio is found to be $\approx 10^{-11}$. The magnetic correction to the correlation
decay (\ref{B.26}) is negligible in this case.

\section{Particle correlations in the many-body system}
We apply the formalism developed in section 3 to the determination of the
large-distance decay of the particle density correlations in the more
general case where all particles are quantum-mechanical, but still obeying
Maxwell-Boltzmann statistics.

We show hereafter that the algebraic $r^{-6}$ decay of the (truncated) particle
density correlations
\begin{align}
    \rho_\text{T}(\gamma_a,\r_a,\gamma_b,\r_b) \sim
    \frac{A_{ab}(\beta,\{\rho_\gamma\})}{|\r_a-\r_b|^{6}}, \qquad |\r_a-\r_b|\to\infty
    \label{particle-correl-decay}
\end{align}
found in the absence of the radiation field \cite{Cornu3},
\cite{Brydges-Martin} is not altered, but that the coefficient
$A_{ab}(\beta,\{\rho_\gamma\})$ contains in addition small magnetic terms
of the order $(\beta m c^2)^{-2}$, as in (\ref{B.26}). As an illustration,
we give the lowest order contribution of this coefficient with respect to
Planck's constant $\hbar$.

By the Feynman-Kac-It\^o representation, the full system composed of
quantum point charges coupled to the radiation field has reduced to a
classical-like system of extended charged loops $\F=(\r,\gamma,\bxi)$ for
which all the methods of classical statistical mechanics apply. The only
novelty comes from the additional magnetic potential $\Wm$.  In the
following, we merely summarize the arguments since they are essentially the
same as those found in \cite{Cornu3}, \cite{Brydges-Martin} when no
radiation field is present.

As usual, we express the truncated two-loop correlation
$\rho_\text{T}(\F_a,\F_b)$ $=$ $\rho(\F_a)\rho(\F_b)h(\F_a,\F_b)$ in terms
of the loop Ursell function $h(\F_a,\F_b)$. The latter function can be
expanded in a formal diagrammatic Mayer series of powers of the loop
densities $\rho(\F)$. One needs to resum the long-range part of the Coulomb
potential $\Vc$, which is responsible for the non-integrability of the
Mayer bonds \\$f(\F_i,\F_j)$ $=$ \mbox{$\exp(-\beta
    e_{\gamma_i}e_{\gamma_j} [\Vc(\F_i,\F_j)+\Wm(\F_i,\F_j)])-1$} at
infinity. Using the decomposition (\ref{B.17}) we resum the convolution
chains built with the purely electrostatic long-range part $\Vel(\F,\F')$
into a Debye-H\"uckel-type screened potential $\Phiel(\F,\F')$. Then
reorganizing the diagrams leads to a representation of the loop Ursell
function by terms of so-called prototype diagrams, built with the two kinds
of bonds
\begin{align}
    &F(\F,\F') = -\beta e_\gamma e_{\gamma'} \Phiel(\F,\F'),
    \\&\FR(\F,\F') = \e^{-\beta e_\gamma e_{\gamma'} [\Phiel(\F,\F') +
            W(\F,\F')]} - 1 + \beta e_\gamma e_{\gamma'} \Phiel(\F,\F'),
\label{FR}
\end{align}
where we have defined $W = \Wc + \Wm$ as in section 4. \footnote{Strictly
    speaking, the short-range repulsive potential needed in the framework
    of Maxwell-Boltzmann statistics would arise here in the exponent of
    (\ref{FR}). It has no implication in this discussion about long-range
    behaviours, and we simply omit it.}

The potential $\Phiel(\F,\F')$ has been studied in
\cite{Ballenegger-etal}. It corresponds to the term $n=0$ of the full
quantum analog of the Debye-H\"uckel potential given by formula (89) of
\cite{Ballenegger-etal}. This contribution $n=0$ is shown to be decaying at
infinity faster than any inverse power of $|\r-\r'|$ (see formula (58) of
\cite{Ballenegger-etal}, and the comment following it).

The asymptotic decay of the two-particle correlation
$\rho_\text{T}(\gamma_a,\r_a,\gamma_b,\r_b)$ is inferred from that of the
loop correlation $\rho_\text{T}(\F_a, \F_b)$ by integrating it over the
Brownian shapes $\bxi_a$ and $\bxi_b$.  The bond $F$ is rapidly decreasing,
and the asymptotic decay of $\FR$ is dominated by the dipolar decays of
$\Wc$ and $\Wm$~: $\FR(\F,\F') \sim -\beta e_\gamma e_{\gamma'} W(\F,\F')$
as $|\r-\r'|\to\infty$. We further extract this dipolar part from $\FR$ and
define the bond
\begin{align}
    \widetilde{\FR}(\F,\F') &= \FR(\F,\F') + \beta e_\gamma e_{\gamma'}
    W(\F,\F') \nonumber
    \\&\sim \tfrac{1}{2}[\beta e_\gamma e_{\gamma'} W(\F,\F')]^2
    =\Order({|\r-\r'|^{-6}})
\label{FRtilde}
\end{align}
and work now with the three bonds $F$, $\widetilde{\FR}$, and
$W$. \footnote{In \cite{Brydges-Martin}, the bond $F$ is further decomposed
    into a multipole expansion. Our bonds $\FR$ and $\widetilde{\FR}$
    differ formally from their bonds $F_\text{l}$ and $\tilde{F}_\text{l}$
    only by the inclusion of the magnetic contribution $\Wm$ into $W$.}

To find out the slowest-decaying diagrams, we write the truncated two-loop
correlation $\rho_\text{T}(\F_a,\F_b)$ in an exact Dyson series of
convolution chains involving $W$ and $H$~:
\begin{align}
	\rho_\text{T}(\F_a,\F_b) = &\rho(\F_a)\rho(\F_b)H(\F_a,\F_b) -\beta (K \star W
    \star K)(\F_a,\F_b) \nonumber
    \\&+ \beta^2 (K \star W \star K \star W \star K)(\F_a,\F_b)+...
    \label{KWseries}
\end{align}
where $H$ denotes the sum of the diagrams that remain connected under
removal of one $W$-bond and
$K(\F_1,\F_2)=\rho(\F_1)\rho(\F_2)H(\F_1,\F_2)+\delta(\F_1,\F_2)\rho(\F_1)$. This
topological constraint ensures that $H$ decays at least as $r^{-6}$. The
series (\ref{KWseries}) is conveniently analysed in Fourier representation
with respect to $\r_a-\r_b$. After expanding $W$ into the sum $\Wc+\Wm$, we
have three types of chains~: pure $\Wc$ or $\Wm$ chains and mixed $\Wc,
\Wm$ chains. It is shown in \cite{Cornu3}, \cite{Brydges-Martin} that the
contribution of pure $\Wc$ chains to the particle correlation
$\rho_\text{T}(\gamma_a,\r_a,\gamma_b,\r_b)=\int \!\D(\bxi_a)
\int\!\D(\bxi_b)\, \rho_\text{T}(\F_a,\F_b)$ decays strictly faster than
$\order(|\r_a-\r_b|^{-6})$. \footnote{In this proof, only the invariance of
    $H$ under rotations is used, which also holds when the magnetic
    potential is included.} We show below that all other chains containing
$\Wm$ bonds vanish identically as the consequence of transversality. This
implies that the longest-range part of the correlations originates from the
function $H$ in the first term of the right-hand side of (\ref{KWseries}),
hence the result (\ref{particle-correl-decay}).

A chain mixing $\Wc$ and $\Wm$ bonds must have at least one element $\Wc
\star K \star \Wm$ or $\Wm \star K \star \Wc$. In Fourier space, one can
write, from (\ref{B.19}) and (\ref{B.14}),
\begin{align}
    (\Wc \star K \star \Wm)&(\gamma_a, \bxi_a, \gamma_b,\bxi_b,\k) =
    \int_0^1\!\!\!\!\d s_a \!\! \int_0^1\!\!\!\!\d s_1\, (\delta(s_a-s_1)-1)
    \frac{4\pi}{k^2} \e^{i \k\cdot\lambda_{\gamma_a} \bxi_a(s_a)}\notag
    \\&\times \big[T^{\nu_2} (\k,s_1) G^{\nu_2,\nu_b}(\k)\big] \int_0^1\!\!\!
    d\xi_b^{\nu_b}(s_b)
    \e^{-i\k\cdot\lambda_{\gamma_b}\bxi_b(s_b)}, \label{WcKWm}
\end{align}
where
\begin{align}
    T^{\nu_2}(\k,s_1) =
    &\sum_{\gamma_1}\!\!\int\!\!\D(\bxi_1)\!\sum_{\gamma_2}\!\!\int\!\!\D(\bxi_2)
    \,\e^{-i\k\cdot\lambda_{\gamma_1}\bxi_1(s_1)} K(\gamma_1, \bxi_1,
    \gamma_2,\bxi_2,\k) \nonumber
    \\& \times\int_0^1
    \!\!\!\d\xi_2^{\nu_2}(s_2) \e^{i\k\cdot\lambda_{\gamma_2}\bxi_2(s_2)}
\end{align}
and $K(\gamma_1, \bxi_1, \gamma_2,\bxi_2,\k)$ is the Fourier transform of
$K(\F_1,\F_2)$ with respect to $\r_1-\r_2$. As the measures $\D(\bxi_1)$
and $\D(\bxi_2)$ and the function $K(\gamma_1, \bxi_1, \gamma_2,\bxi_2,\k)$
are invariant under spatial rotations, $T^{\nu_2}(\k,s_1)$ transforms as a
tensor, implying that it is necessarily of the form $T^{\nu_2}(\k,s_1) =
k^{\nu_2}\, a(k,s_1)$ for some rotationally invariant function $a$ of
$\k$. Using $k^\mu G^{\mu\nu} (\k) \equiv 0$ one deduces immediately that
(\ref{WcKWm}) vanishes. The case of $\Wm\star K\star \Wc$ is similar. To
see that there are no chains containing only $\Wm$ bonds in
$\rho_\text{T}(\gamma_a,\r_a,\gamma_b,\r_b)$, it is sufficient to notice
that the integrated root element $\int\!  \D(\bxi_a)\, K \star \Wm$ also
involves a factor $[T^{\nu_2}(\k) G^{\nu_2,\nu_b}(\k)]$ (for another
function $T^{\nu_2}(\k)$ transforming in a covariant manner), and thereby
vanishes for the same reason.

The graphs that do contribute to the coefficient
$A_{ab}(\beta,\{\rho_\gamma\})$ of (\ref{particle-correl-decay}) are those
of $H$ that contain bonds with algebraic decay~: namely, $\widetilde{\FR}$
and $W$. To select the lowest contribution in $\hbar$, one notes first that
$W$ is at least of order $\hbar^2$, as seen in (\ref{B.16}), (\ref{B.21})
which correspond to the lowest-order terms in the multipolar expansions of
$\Wc$ and $\Wm$. (Higher-order multipoles generate higher powers of the de
Broglie wavelengths.) Since $\Phiel$ is rapidly decreasing, the algebraic
part of $\widetilde{\FR}$ is of order $\hbar^4$ and is given by
$\tfrac{1}{2}[\beta e_\gamma e_{\gamma'} W(\F,\F')]^2$, as in
(\ref{FRtilde}). Thus, up to order $\hbar^4$, graphs with an algebraic
decay can contain only one bond $W$, two bonds $W$, or one bond
$\widetilde{\FR}$ belonging to paths connecting the two root points. If
there is a single such link $W$, by the topological structure of $H$ there
exists another path connecting the root points made of the more rapidly
decreasing bonds $F$ and $\widetilde{\FR}$. Hence the whole graph has a
decay faster than $r^{-6}$. If there are two $W$ bonds in between the root
points, as each of them is of order $\hbar^2$ all the other bonds and
vertices can be evaluated in the classical limit $\hbar\to
0$. Consequently, at least one of the extremities of either bond $W$ is
attached to a purely classical part of the graph, which is independent of
the Brownian shapes. We call such a point a classical end of $W$. At such
points, integration over the Brownian shape of the loop ``kills'' the
$r^{-3}$ decay of $W$ (see Appendix C), leading to an overall decay faster
than $r^{-6}$. Finally, at order $\hbar^4$, the only graphs that contribute
to (\ref{particle-correl-decay}) are constituted by a single
$\widetilde{\FR}$ bond linked to the root points by purely classical
subgraphs. The sum of such graphs contributes to the particle correlation
in the large-distance limit as
\begin{align}
	\rho_\text{T}(\gamma_a,\r_a,\gamma_b,\r_b) \sim &\sum_{\gamma_1,\gamma_2} \left[
    \int\!\! \d \r\ n_\text{T}^\text{cl}(\gamma_a,\gamma_1,\r)\right] \left[ \int\!\! \d \r
    \ n_\text{T}^\text{cl}(\gamma_2,\gamma_b,\r)\right] \label{particle-correl-asympt-graph}
    \\&\times \int\!\! \D(\bxi_1) \!\!\int\!\! \D(\bxi_2)\ \tfrac{1}{2} \left[
        \beta e_{\gamma_1} e_{\gamma_2} W^\text{dip}(\gamma_1,\bxi_1,\gamma_2,
        \bxi_2, \r_a-\r_b)\right]^2, \nonumber
\end{align}
where $W^\text{dip}=\Wc^\text{dip} + \Wm^\text{dip}$ is the sum of the
dipolar parts (\ref{B.21}) and (\ref{B.16}) of $\Wc$ and $\Wm$, and
$n_\text{T}^\text{cl}(\gamma_a,\gamma_1,\r)$ is the classical truncated
two-point density correlation (including coincident points). The functional
integrals in (\ref{particle-correl-asympt-graph}) have been calculated in
section 4, see (\ref{B.28})-(\ref{B.26}), yielding the final result
\begin{align}
	\rho_\text{T}(\gamma_a,\r_a,\gamma_b,\r_b) \sim &\frac{\hbar^4
        \beta^4}{240}\sum_{\gamma_1,\gamma_2} \left[ \int\!\! \d
    \r\ n_\text{T}^\text{cl}(\gamma_a,\gamma_1,\r)\right] \left[ \int\!\! \d \r
    \ n_\text{T}^\text{cl}(\gamma_2,\gamma_b,\r)\right]\nonumber\\
	&\times \frac{e_{\gamma_1}^2 e_{\gamma_2}^2}{m_{\gamma_1} m_{\gamma_2}}
    \left[ 1 + \frac{5}{\beta m_{\gamma_1} c^2 \beta m_{\gamma_2} c^2}\right]
    \frac{1}{|\r_a-\r_b|^{6}}
\end{align}
as $|\r_a-\r_b|\to\infty$ and at lowest order in $\hbar$. To this order, the
only difference with (\ref{B.26}) is the occurrence of the classical correlation
functions $n_\text{T}^\text{cl}$, a manifestation of the fact that in the quantum many-body
 problem, every pair of particles contribute to the tail of the
correlation function. This generalizes the result of
\cite{Alastuey-Martin}, formula (5.12), to the inclusion of the magnetic
interactions.

As a final comment, we observe that the inclusion of the transverse degrees of
freedom of the field does not modify the charge sum rule in the system of loops
and hence it also holds for the charge correlations in the particle system. This
sum rule reads
\begin{align}
	\int\!\!\d\r \!\int\!\! \D(\bxi) \sum_{\gamma} \frac{e_{\gamma} \rho_\text{T}(\F,
        \F_1)}{\rho(\F_1)} = -e_{\gamma_1}.
\end{align}
It states that the charge of the cloud of loops induced around a fixed loop
$\F_1$ exactly compensates that of $\F_1$. The proof can be carried out word by
word as in \cite{Ballenegger-etal}, section 6.1.2. It relies exclusively on the
long-range part $r^{-1}$ of the Coulomb potential $\Vc$ and is not altered by
the presence of the magnetic potential $\Wm$.

\section{Transverse field correlations}

A characteristic feature of charged systems is that longitudinal field
correlations always remain long ranged in spite of the screening mechanisms
that reduce the range of the particle correlations. It has been established
on a microscopic basis that the correlations of the longitudinal electric
field $\mathbf{E}_\text{l}$ behave as \cite{Lebowitz-Martin},
\cite{martin-sumrules}
\begin{align}
    \langle \EL^\mu(\bx)\EL^\nu(\by)\rangle_\text{T} \sim -\partial_\mu
    \partial_\nu \frac{1}{|\bx-\by|} \left[-\tfrac{2\pi}{3}\int\!
        \d\r\ |\r|^{2} S(\r)\right] , \quad |\bx-\by|\to \infty, \la{F.1}
\end{align}
where $S(\r)$ is the (classical or quantum-mechanical) charge-charge
correlation function.


In order to obtain the correlations of the transverse fields we first
consider correlations $\langle
A^{\mu}(\mathbf{x})A^{\nu}(\mathbf{y})\rangle_\text{T}$ of the vector
potential at free points $\bx$ and $\by$ in space.  These correlations are
easily obtained by functional differentiation, adding to the original
Hamiltonian (\ref{B.1}) a coupling to an external current
$\bJ_\text{ext}(\bx)$
\begin{align}
    H_{L,R}(\bJ_\text{ext})=H_{L,R}-i\! \int\!\!
    \d\mathbf{x}\ \bJ_\text{ext}(\mathbf{x})\cdot\mathbf{A}(\mathbf{x}), \la{F.2}
\end{align}
so that
\begin{align}
    \langle A^{\mu}(\mathbf{x})A^{\nu}(\mathbf{y})\rangle_\text{T}=\left.
    -\frac{1}{\beta^{2}} \frac {\delta^{2}}
    {\delta\mathcal{J}_\text{ext}^{\mu}(\mathbf{x})
        \delta\mathcal{J}_\text{ext}^{\nu}(\mathbf{y})}\ln \Tr\ \e^{-\beta
        H_{L,R}(\bJ_\text{ext})} \right|_{\bJ_\text{ext}=0}.  \la{F.3}
\end{align}
Decomposing $H_{L,R}$ as in (\ref{B.4b}) one can write
\begin{align}
    &\langle A^{\mu}(\mathbf{x})A^{\nu}(\mathbf{y})\rangle_\text{T} = \nonumber\\
    &=\left.  -\frac{1}{\beta^{2}} \frac {\delta^{2}}
    {\delta\mathcal{J}_\text{ext}^{\mu}(\mathbf{x})\delta
        \mathcal{J}_\text{ext}^{\nu}(\mathbf{y})}\ln \Tr_\text{mat} \left\langle
    \e^{-\beta H_\mathbf{A}}\e^{i\beta \int\! \d\mathbf{x}\ \bJ_\text{ext}(\mathbf{x})\cdot
        \mathbf{A}(\mathbf{x})}
    \right\rangle_\text{rad}\right|_{\bJ_\text{ext}=0}.
\la{F.4}
\end{align}
Using the Feynman-Kac formula as in section 3 one sees that the only
modification in (\ref{B.9}) is the replacement of the loop current
$\bJ(\bx)$ by the total current\footnote{As a consequence of the imaginary
    coupling constant in the Hamiltonian (\ref{F.2}), the total current is
    real, so that we can still apply the Gaussian integration formula used
    in (\ref{B.9}).}
\begin{align}
    \bJ_\text{tot}(\bx)=\bJ(\bx)+\bJ_\text{ext}(\bx). \la{F.5}
\end{align}
The Gaussian integration on the field variables replaces (\ref{B.9})
by
\begin{align}
    &\exp\Big\{-\frac{\beta}{2} \int\!\!\! \frac{\d\mathbf{k}}{(2\pi)^{3}}
    \left(\mathcal{J}_\text{tot}^{\mu}(\mathbf{k})\right)^{*}
    G^{\mu\nu}(\mathbf{k})\mathcal{J}_\text{tot}^{\nu}(\mathbf{k}) \Big\}=
    \exp\Big\{-\frac{\beta}{2} \int\!\!\! \frac{\d\mathbf{k}}{(2\pi)^{3}}
    G^{\mu\nu}(\mathbf{k})
    \nonumber\\
    &\times\left[\left(\mathcal{J}^{\mu}\right)^{*}
        \mathcal{J}^{\nu} +
        \left(\mathcal{J}_\text{ext}^{\mu}\right)^{*}
        \mathcal{J}^{\nu} +
        \left(\mathcal{J}^{\mu}\right)^{*}
        \mathcal{J}_\text{ext}^{\nu} +
        \left(\mathcal{J}_\text{ext}^{\mu}\right)^{*}
        \mathcal{J}_\text{ext}^{\nu} \right](\k)\Big\}.
    \la{F.6}
\end{align}
Therefore, from (\ref{F.6}), functional differentiation with
respect to $\bJ_\text{ext}$ according to (\ref{F.4}) produces two terms
\begin{align}
    \langle A^{\mu}(\mathbf{x})A^{\nu}(\mathbf{y})\rangle_\text{T} =
    \langle A^{\mu}(\mathbf{x})A^{\nu}(\mathbf{y})\rangle^{0}_\text{T} + \langle A^{\mu}(\mathbf{x})
    A^{\nu}(\mathbf{y})\rangle^\text{mat}_\text{T}.
\la{F.7}
\end{align}
The first contribution (written in Fourier form)
\begin{align}
    \langle A^{\mu}(\mathbf{x})A^{\nu}(\mathbf{y})\rangle^{0}_\text{T} &=
    \frac{1}{\beta}\int\!\!\! \frac{\d\mathbf{k}}{(2\pi)^{3}}
    \e^{i\mathbf{k}\cdot (\mathbf{x}-\mathbf{y})} G^{\mu\nu}(\mathbf{k})\nonumber\\
    &\sim \frac{1}{2\beta r}\(\delta^{\mu\nu} + \frac{r^{\mu}r^{\nu}}{r^{2}}\),
    \quad r\to\infty, \;\;\r=\bx-\by,
\la{F.8}
\end{align}
arises from the part quadratic in $\bJ_\text{ext}$ in (\ref{F.6}). It
describes the thermal fluctuations of the free field, and in view of
(\ref{B.12}), decays as $r^{-1}$.  The second term, coming from the part
linear in $\bJ_\text{ext}$,
\begin{align}
    &\langle A^{\mu}(\mathbf{x})A^{\nu}(\mathbf{y})\rangle^\text{mat}_\text{T}=\nonumber\\
    &=-\int\!\!\! \frac{\d\mathbf{k}}{(2\pi)^{3}}\, \e^{i\mathbf{k\cdot x}}
    \!\!\int\!\!\! \frac{\d\mathbf{k}'}{(2\pi)^{3}}\, \e^{i\mathbf{k}'\cdot\mathbf{y}}
    G^{\mu\sigma}(\mathbf{k}) G^{\nu\tau}(\mathbf{k}')
    \ \langle\mathcal{J}^{\sigma}(\mathbf{k})
    \mathcal{J}^{\tau}(\mathbf{k}')\rangle_\text{T}, \la{F.9}
\end{align}
represents the modification to the free-field fluctuations caused by the
presence of matter. It involves the loop current correlation function
$\langle\mathcal{J}^{\sigma}(\mathbf{k}) \mathcal{J}^{\tau}(\mathbf{k}')
\rangle_\text{T}$ where the average is taken with respect to the thermal
weight (\ref{loop-Gibbs}) for the statistical-mechanical system of
loops. Expressing the currents $\pmb{\mathcal{J}}(\k) = \int \d
\F\ \mathbf{j}(\F,\k) \hat\rho(\F)$ in terms of the density of loops
$\hat\rho(\F)=\sum_i \delta(\F,\F_i)$ (see (\ref{B.8})), one can write this
current correlation in terms of the loop density correlation function
$n_\text{T}(\gamma_{1},\bxi_{1},\gamma_{2},\bxi_{2},\mathbf{k})$ (including
the contribution of coincident points)~:
\begin{align}
    &\langle\mathcal{J}^{\sigma}(\mathbf{k})\mathcal{J}^{\tau}(\mathbf{k}')\rangle_\text{T} =
    (2\pi)^{3}\delta(\mathbf{k} + \mathbf{k}')\nonumber\\
    &\times\sum_{\gamma_{1}, \gamma_{2}} \!\int\!\! \D(\bxi_{1})
    \!\!\int\!\! \D(\bxi_{2})\ \mathcal{T}^{\sigma}(\gamma_{1},
    \bxi_{1}, \mathbf{k}) \big(\mathcal{T}^{\tau}(\gamma_{2},
    \bxi_{2}, \mathbf{k})\big)^\ast n_\text{T}(\gamma_{1}, \bxi_{1}, \gamma_{2},
    \bxi_{2}, \mathbf{k}).
\la{F.10}
\end{align}
The $\delta(\mathbf{k}+\mathbf{k}')$ is the manifestation of the translational
invariance of the state, and we have set
\begin{align}
    \mathcal{T}^{\sigma}(\gamma_{i}, \bxi_{i}, \mathbf{k}) =
    \frac{e_{\gamma_{i}}}{\sqrt{\beta m_{\gamma_{i}}c^{2}}} \int_{0}^{1}\!\!\!
    \d\xi_{i}^{\sigma}(s_{i})\ \e^{i\lambda_{\gamma_{i}} \mathbf{k} \cdot\bxi_{i}(s_{i})}.
\la{F.11}
\end{align}
When (\ref{F.10}) is introduced into (\ref{F.9}), one obtains the final form
\begin{align}
    \langle A^{\mu}(\mathbf{x})A^{\nu}(\mathbf{y})\rangle_\text{T}^\text{mat}=
    -\int\!\!\!\frac{\d\mathbf{k}}{(2\pi)^{3}}\, \e^{i\bk\cdot (\bx-\by)}
G^{\mu\sigma}(\mathbf{k})G^{\nu\tau}(\mathbf{k})\mathcal{Q}^{\sigma\tau}(\mathbf{k}),
    \la{F.12}
\end{align}
where $\mathcal{Q}^{\sigma\tau}(\mathbf{k})$ is the tensor
\begin{align}
    \mathcal{Q}^{\sigma\tau}(\mathbf{k}) = \sum_{\gamma_{1}, \gamma_{2}}
    \!\int\!\! \D(\bxi_{1}) \!\!\int\!\! \D(\bxi_{2})\,
    \mathcal{T}^{\sigma}(\gamma_{1}, \bxi_{1}, \mathbf{k})
    \big(\mathcal{T}^{\tau}(\gamma_{2}, \bxi_{2}, \mathbf{k})\big)^\ast
    n_\text{T}(\gamma_{1}, \bxi_{1}, \gamma_{2}, \bxi_{2}, \mathbf{k}).
    \la{F.13}
\end{align}
To obtain the long-distance behaviour of this correlation we examine the
integrand in (\ref{F.13}) at small $\bk$. Because of isotropy, the tensor
$\mathcal{Q}^{\sigma\tau}(\mathbf{k})$ transforms covariantly under the
rotations, and thus is of the form
\begin{align}
    \mathcal{Q}^{\sigma\tau}(\mathbf{k})=a(k)\delta^{\sigma\tau}+b(k)
    k^{\sigma}k^{\tau}, \quad k=|\bk|
\la{F.14}
\end{align}
The term $k^{\sigma}k^{\tau}$ does not contribute to (\ref{F.13}) since
$G^{\mu\sigma}(\mathbf{k})$ is transversal. Because of It\^o's lemma,
$\mathcal{T}^{\sigma}(\gamma_{i},\bxi_{i},\mathbf{k})$ is linear in $\bk$
as $\bk\to 0$, implying $a(k)=a\;k^{2}[1 + o(k)]$.  Hence, using
$\delta^{\mu\sigma}_{tr}(\bk)\delta^{\nu\sigma}_{tr}(\bk)=\delta^{\mu\nu}_{tr}(\bk)$
one finds
\begin{align}
G^{\mu\sigma}(\mathbf{k})G^{\nu\tau}(\mathbf{k})\mathcal{Q}^{\sigma\tau}(\mathbf{k})=4\pi
    \,a\, \frac{4\pi}{k^{2}}\,\delta^{\mu\nu}_{tr}(\bk) [1+o(k)]=4\pi \,a\,
    G^{\mu\nu}(\bk)[1+o(k)] \la{F.15}
\end{align}
as $\ k\to 0$. This shows that $\langle
A^{\mu}(\mathbf{x})A^{\nu}(\mathbf{y})\rangle_\text{T}^\text{mat}$ has the
same type of decay as the free field part (\ref{F.8}) with a modified
amplitude. Summing up the two contributions (\ref{F.7}) gives
\begin{align}
    \langle A^{\mu}(\mathbf{x})A^{\nu}(\mathbf{y})\rangle_\text{T}\sim
\frac{1}{2r}\left(\delta^{\mu\nu}+\frac{r^{\mu}r^{\nu}}{r^{2}}\right)
    \left(\frac{1}{\beta}-4\pi a\right), \quad r\to\infty. \la{F.16}
\end{align}
For $\mathbf{B}(\bx)=\nabla\times\A(\bx)$, one finds, from (\ref{F.16}),
\begin{align}
	\langle B^{\mu}(\mathbf{x})B^{\nu}(\mathbf{y})\rangle_\text{T}  \sim
\left(\partial_\mu \partial_\nu \frac{1}{r}\right)
    \left(\frac{1}{\beta}-4\pi a\right), \quad r\to\infty. \label{correl-B}
\end{align}
The constant $a=a(\hbar,\beta,\rho)$ embodies the effects of matter on the
transverse field fluctuations. It has a relativistic factor $(m c^2)^{-1}$
and vanishes in the classical limit $\hbar\to 0$ (in accordance to the
Bohr--van Leeuwen decoupling) as $\Order(\hbar^4)$ (see Appendix D).

In order to find the correlations of the transverse electric field
\begin{align}
	\mathbf{E}_\text{t}(\bx)&=-\frac{1}{c}\left.\frac{\partial \A(\bx,t)}{\partial t}\right\vert_{t=0}  \nonumber
	\\&= -\(\frac{4\pi \hbar c^2}{R^{3}}\)^{1/2}\sum_{\bk\lambda}g(\bk)
    \frac{ {\bf e}_{\bk\lambda}}{\sqrt{2\omega_{\bk}}} \Big(\frac{i
        \omega_\k}{c} \alpha_{\bk\lambda}^{*}\e^{-i\bk\cdot\bx}- \frac{i
        \omega_\k}{c} \alpha_{\bk\lambda}\e^{i\bk\cdot\bx}\Big),
\end{align}
we couple the latter to an external polarisation $\bP_\text{ext}(\bx)$,
\begin{align}
    H_{L,R}(\bP_\text{ext})=H_{L,R}-i\! \int\!\!
    \d\mathbf{x}\ \bP_\text{ext}(\mathbf{x})\cdot\mathbf{E}_\text{t}(\mathbf{x}), \la{F.2a}
\end{align}
and proceed as after (\ref{F.2}). This amounts to replacing everywhere
$\bJ_\text{ext}(\k)$ by $i k \bP_\text{ext}(\k)$ so that the right-hand
side of equation (\ref{F.6}) is changed into
\begin{align}
&\exp\Big\{-\frac{\beta}{2} \int\!\!\! \frac{\d\mathbf{k}}{(2\pi)^{3}}
    G^{\mu\nu}(\mathbf{k}) \label{cov-Pext}
    \\
    &\times\left[\left(\mathcal{J}^{\mu}\right)^{*}
        \mathcal{J}^{\nu} - i k
        \left(\mathcal{P}_\text{ext}^{\mu}\right)^{*}
        \mathcal{J}^{\nu} + i k
        \left(\mathcal{J}^{\mu}\right)^{*}
        \mathcal{P}_\text{ext}^{\nu} + k^2
        \left(\mathcal{P}_\text{ext}^{\mu}\right)^{*}
        \mathcal{P}_\text{ext}^{\nu} \right](\k)\Big\}. \nonumber
\end{align}
As $\pmb{\mathcal{P}}_\text{ext}(\r)$ and $\pmb{\mathcal{J}}(\r)$ are real,
$\pmb{\mathcal{P}}_\text{ext}^\ast(\k) = \pmb{\mathcal{P}}_\text{ext}(-\k)$
and likewise for $\pmb{\mathcal{J}}$. From the change of variable
$\k\mapsto -\k$, one sees that the second term in the integrand in
(\ref{cov-Pext}) is exactly compensated by the third term. Only the term
quadratic in $\pmb{\mathcal{P}}_\text{ext}$ remains, which is responsible
upon functional differentiation for the thermal fluctuations of the free
field, as in (\ref{F.8}). Hence, the correlations of the transverse part of
the electric field are decoupled from matter and one finds
\begin{align}
	\langle \ET^{\mu}(\mathbf{x})\ET^{\nu}(\mathbf{y})\rangle_\text{T} =
    \langle\ET^{\mu}(\mathbf{x})\ET^{\nu}(\mathbf{y})\rangle_\text{T}^0
    \sim \left(\partial_\mu \partial_\mu \frac{1}{r}\right)
    \frac{1}{\beta}, \quad r\to\infty. \label{correl-E-trans}
\end{align}
The asymptotic correlation of the complete electric field $\mathbf{E}(\bx)
= \mathbf{E}_\text{l}(\bx) +\mathbf{E}_\text{t}(\bx)$ follows from
(\ref{F.1}) and (\ref{correl-E-trans}) (one can check by similar
calculations that the cross correlation $\langle
\EL^{\mu}(\mathbf{x})\ET^{\nu}(\mathbf{y})\rangle_\text{T}$ vanishes
identically)~:
\begin{align}
	\langle E^{\mu}(\mathbf{x})E^{\nu}(\mathbf{y})\rangle_\text{T} &=
    \langle \EL^{\mu}(\mathbf{x})\EL^{\nu}(\mathbf{y})\rangle_\text{T}
    +\langle \ET^{\mu}(\mathbf{x})\ET^{\nu}(\mathbf{y})\rangle_\text{T}
    \nonumber
	\\&=\left(\partial_\mu \partial_\mu \frac{1}{r}\right)
    \left(\tfrac{2\pi}{3}\int\! \d\r\ |\r|^{2}
    S(\r)+\frac{1}{\beta}\right), \quad r\to\infty. \label{correl-E}
\end{align}
In the classical limit, $S(\r)$ satisfies the second-moment
Stillinger--Lovett sum rule \cite{martin-sumrules} $-\tfrac{2\pi}{3}\int\!
\d\r\ |\r|^{2} S(\r) = 1/\beta$. Hence, the asymptotic longitudinal
electric field correlations in the matter are exactly compensated by those
of the free radiation part, as noted in \cite{Felderhof}. However, this no
longer holds for quantum plasmas. As an illustration, for the quantum
one-component plasma (jellium), one has \cite{pines-nozieres}
\begin{align}
-\tfrac{2\pi}{3}\int\! \d\r\ |\r|^{2} S(\r) = \tfrac{\hbar \omega_p}{2}
\text{coth}\left(\frac{\hbar \omega_p \beta}{2}\right) = \frac{1}{\beta} +
\frac{\beta}{3} \left(\frac{\hbar \omega_p}{2}\right)^2
+\Order(\hbar^4), \label{ocp-sumrule}
\end{align}
where $\omega_p$ is the plasmon frequency. The long range of the electric
field correlations is thus a purely quantum-mechanical effect.  These
findings are further discussed in the concluding remarks (section $8$).

\section{Bose and Fermi statistics}
In this final section we introduce the needed modifications when the
Fermionic or Bosonic particle statistics are taken into account.

The Bose or Fermi statistics of the particles can be incorporated in the
formalism following the same procedure as described in \cite{Cornu1},
\cite{Brydges-Martin} (section V). The matrix elements of (\ref{B.5}),
which is still an operator depending on the particle variables, are to be
evaluated with properly symmetrized (antisymmetrized) states. When
combining the decomposition of the permutation into cycles with the
Feynman-Kac-It\^o path integral representation this leads to generalize the
previous loops $\F=(\r, \gamma, \b\xi)$ to Brownian loops
$\cl=(q,\bR,\gamma,\bX)$ that carry $q$ particles (a set of particles
labeled by indices belonging to a permutation cycle of length $q$).  The
loop is located at $\bR$ and has a random shape which is a Brownian bridge
$\bX(s)$, $0\leq s\leq q$, $\bX(0)=\bX(q)=\mathbf{0}$ with zero mean and
covariance
\begin{align}
    \int\!\! \D({\bf X })\, X_{\mu} (s_{1})
    X_{\nu}(s_{2})=\delta_{\mu\nu}\,q\left[\min\left(\frac{s_{1}}{q},\frac{s_{2}}{q}\right)
        -\frac{s_{1}}{q}\frac{s_{2}}{q}\right].
\la{7.14}
\end{align}
We merely give the final formulae since all steps are essentially identical
as those presented in the above mentioned works.

The grand canonical partition function of the particle system, with the
field degrees of freedom integrated out, takes the following classical-like
form in the space of loops
\begin{align}
    \Xi_\Lambda=\sum_{n=0}^\infty\frac{1}{n!}\int\!\prod_{i=1}^n\d\cl_i\ z(\cl_i)
    \exp\big(-\beta \mathcal{U}(\cl_1,\ldots,\cl_n)\big).
\label{7.15}
\end{align}
Integration on phase space means integration over space and summation over
all internal degrees of freedom of the loops~:
\begin{align}
    \int\!\! \d\cl\cdots=\int\!\! \d\bR \sum_{\gamma}\sum_{q=1}^\infty \int\!\!
    \D(\bX)\cdots.
\label{7.16}
\end{align}
$\mathcal{U}(\cl_1,\ldots,\cl_n)$ is the sum of the pair interactions
between two different loops
$e_{\gamma_i}e_{\gamma_j}[\mathcal{V}_\text{c}(\cl_i,\cl_j)+\mathcal{W}_\text{m}(\cl_i,\cl_j)]$
with
\begin{align}
    \mathcal{V}_\text{c}(\cl_i,\cl_j) =\int_0^{q_{i}}\!\!\! \d s_{1}
    \!\!\int_0^{q_{j}}\!\!\! \d s_{2}\ \delta( \tilde{s}_1-\tilde{s}_2)
    \frac{1}{\big|\bR_i+\lambda_{\gamma_i}\bX_{i}(s_1) -
        \bR_j-\lambda_{\gamma_j}\bX_{j}(s_2)\big|} \la{7.17}
\end{align}
the Coulomb potential, and
\begin{align}
    \mathcal{W}_\text{m}(\cl_i,\cl_j)&=\frac{1}{\beta
        \sqrt{m_{\gamma_i}m_{\gamma_j}}c^{2}}\int\!\!\! \frac{\d\bk }{(2\pi)^{3}}\,
    \e^{i\bk\cdot(\r_i-\r_j)}\la{7.18}\\
    &\times\int_{0}^{q_{i}}\!\!\! \d X_{i}^\mu(s_1)\,
    \e^{i\bk\cdot\lambda_{\gamma_{i}} \bX_{i}(s_1)}\! \int_{0}^{q_{j}}\!\!\! \d
    X_{j}^\nu(s_2)\,
    \e^{-i\bk\cdot\lambda_{\gamma_{j}}\bX_{j}(s_2)}\
   G^{\mu\nu}(\bk) \nonumber
\end{align}
the effective magnetic potential obtained after integrating out the field
variables. Here ${\tilde s}=s\ \text{mod}\ 1$ and
$
    \delta({\tilde s})=\sum_{n=-\infty}^{\infty}\e^{2i\pi ns} 
$
is the periodic Dirac function of period $1$ that takes into account the
equal time constraint imposed by the Feynman-Kac formula.  Finally, the
activity $z(\cl_i)$ of a loop
\begin{align}
    z(\cl_{i})=\frac{(\eta_{\gamma_i})^{q_{i}-1}}{q_{i}}\;
    \frac{z_{\gamma_{i}}^{q_i}}{(2\pi q_i \lambda_{\gamma_{i}}^2)^{3/2}}
    \;\exp(-\beta [\mathcal{U}_\text{self}(\cl_{i})+\mathcal{V}_\text{walls}(\cl_i)]), \;\;\;z_{\gamma_{i}}=\e^{\beta
        \mu_{\gamma_{i}}}
\label{7.20} 
\end{align}
incorporates the chemical potential $\mu_{\gamma_{i}}$ of the particle, the
effects of quantum statistics ($\eta_{\gamma_{i}}=1$ for bosons and
$\eta_{\gamma_{i}}=-1$ for fermions), and the internal interaction
$\mathcal{U}_\text{self}(\cl_{i})=-\frac{\beta
    e_i^2}{2}(\mathcal{V}_\text{c}+\mathcal{W}_\text{m})(\cl_i,\cl_i)$ of
the particles belonging to the same loop (omitting the infinite Coulomb
self-energies of the point particles). The addition of the magnetic
potential $\mathcal{W}_\text{m}$ is the only modification compared to the
formalism previously developed for pure Coulombic
interactions. Maxwell-Boltzmann statistics and the potentials (\ref{B.14})
and (\ref{B.6}) of section 3 are recovered when only single-particle loops
($q=1$) are retained.

At this point, due to the classical-like structure of the partition
function (\ref{7.15}), methods of classical statistical mechanics can be
used in the auxiliary system of loops, in particular the technique of Mayer
graphs, as in section 5. The statistics of the particles affects the
coefficients of the tails of the density and field correlations, but not
their general forms (\ref{particle-correl-decay}), (\ref{correl-B}) and
(\ref{correl-E}).


\section{Concluding remarks}
The Feynman--Kac--It\^o path integral representation of the Gibbs weight
enables one to integrate out the (classical) field variables. This yields
an exact pairwise magnetic potential in the space of loops, which is
asymptotically dipolar. It generates small ($\Order((\beta m c^2)^{-2})$)
corrections to the tail of the particle correlation due to the pure
Coulombic interaction.

A word is necessary about spin interactions that have not been included in
the above discussion. Spin-$1/2$ electrons give rise to the additional term
$-\nu\sum_{i=1}^{n}\bs_{i}\cdot \bB(\r_{i})$ in the Hamiltonian, with
$\bB(\r)=\nabla \wedge \A(\r)$, $\nu=\tfrac{g_{s}e\hbar}{4mc}$, $g_{s}$ the
gyromagnetic factor, and $\bs$ the Pauli matrices. Using spin coherent
states \cite{klauder}, a functional integral representation of the Gibbs
weight can be constructed including the coupling of the spins to the
field. Since this coupling is linear with respect to the vector potential,
Gaussian integration on the field variables leads again to an effective
spin-spin interaction $W_\text{s}(i,j)$ and effective cross interactions
$W_\text{sm}(i,j)$ and $W_\text{ms}(i,j)$ between spin and orbital
magnetism; details can be found in \cite{Diplome}. One finds that these
interactions are of dipolar type $\sim r^{-3}, r\to\infty$ and they have to
be added to the magnetic potential $\Wm(i,j)$. In a homogeneous and
isotropic phase, the spin interaction terms contribute to the $r^{-6}$ tail
of the particle correlations with the same amplitude
$\frac{\lambda_{a}^{2}\lambda_{b}^{2}e_{a}^{2}e_{b}^{2}\rho_{a}\rho_{b}}{m_{a}m_{b}c^{4}}$,
up to numerical factors, as that found in section 4 in the case of the
magnetic potential $\Wm$.

Regarding the electric field correlations in the plasma, we also find that
they have an algebraic decay of dipolar type. This is in disagreement with
the macroscopic calculation presented by Landau and Lifshitz \cite{Landau},
\S $88$, based on linear response theory and the fluctuation-dissipation
theorem. Indeed, the electric field fluctuations obtained in this theory
are short ranged (exponentially fast decaying)~: unlike in our calculation,
the algebraic parts of the longitudinal and transverse correlations
compensate exactly in the Landau and Lifshitz formulae
\cite{jancovici-private-comm}. Understanding the relation between our
strictly microscopic approach and the macroscopic theory of field
fluctuations is an open problem.

Let us, however, briefly point out some differences between the two
approaches. The microscopic approach involves all length scales, whereas
Landau and Lifshitz assume that the correlations of the polarisation are
local ($\delta$ correlated in space) and thus deal with a
wave-number-independent dielectric function $\epsilon(\omega)$. Taking into
account the wave-number dependence, it is likely that $\epsilon(\k, \omega)$
has terms non-analytic in $\k$, reflecting the fact that Coulombic matter
has algebraically decaying correlations. In fact, for the jellium model,
the static dielectric function $\epsilon(\k,\omega=0)$ has a singular term
$\sim |\k|$ in its small-$\k$ expansion \cite{Cornu-Martin}. It is possible
that in a non-local generalization of the Landau--Lifshitz theory such
singular terms also generate power-law decays of the field
correlations. Furthermore, the magnetic permeability is usually set equal
to that of the vacuum, thus ignoring the magnetization fluctuations,
whereas in our calculation the latter are properly included.

We stress again that the results of this paper hold when the
electromagnetic field is classical, which is supposed to correctly depict
the small-wave-number regime, as said in the Introduction. Hence, the
occurrence of the Planck constant originates exclusively from the
quantum-mechanical nature of matter. If the field is quantized, we can,
however, not exclude an interplay between $\hbar_\text{matter}$ and
$\hbar_\text{field}$, which could lead to a modification of the asymptotic
formulae presented in the paper.

\subsection*{Acknowledgements}
We thank A. Alastuey and B. Jancovici for useful discussions. P.R.B. is
supported by the Swiss National Foundation for Scientific Research.

\appendix

\section*{Appendix A}

To establish (\ref{B.31}) according to the middle point prescription
(\ref{3.3}) one has to evaluate the rotationally covariant tensor
\begin{align}
    &\int\!\!\D(\bxi) \int_{0}^{1}\!\!\! \d\xi^{\alpha}(s)
    \!\int_{0}^{1}\!\!\!  \d\xi^{\gamma}(t)\ \xi^{\omega}(s)\xi^{\epsilon}(t) =
    \la{A.1}\\
    & = \lim_{n,m\rightarrow\infty} \sum_{k,l=1}^{n,m}
    \int\!\! \D(\bxi) \left[\xi^{\alpha} \left(k_{n}\right) \!-\!
    \xi^{\alpha} \left(k_{n}\!-\!\tfrac{1}{n} \right) \right] \left[\xi^{\gamma}
    \left(l_{m}\right) \!-\! \xi^{\gamma} \left(l_{m}\!-\!\tfrac{1}{m} \right)
    \right]\nonumber\\
    &\times \frac{1}{2} \left[\xi^{\omega}\left(k_{n}\right) \!+\! \xi^{\omega}
    \left(k_{n} \!-\! \tfrac{1}{n}\right)\right] \frac{1}{2} \left[\xi^{\epsilon}
    \left(l_{m}\right) \!+\! \xi^{\epsilon}\left(l_{m}\!-\!\tfrac{1}{m}\right)\right]=
    \delta^{\alpha\gamma}\delta^{\omega\epsilon}A_1+
    \delta^{\alpha\omega}\delta^{\gamma\epsilon}A_2+
    \delta^{\alpha\epsilon}\delta^{\gamma\omega}A_3
\nonumber,
\end{align}
where $k_{n}=\tfrac{k}{n}$ and $l_{m}=\tfrac{l}{m}$.
Setting $C(s,t) = \delta^{\mu\nu}(\min(s,\:t)-s t)$ (see (\ref{3.2})), one has
\begin{align}
    A_1 &=\lim_{n,m\rightarrow\infty}\frac{1}{4}\left[C\(k_{n},l_{m}\) -
        C\(k_{n},l_{m} \!-\! \tfrac{1}{m}\)-C\(k_{n} \!-\! \tfrac{1}{n},l_{m}\) +
        C\(k_{n}\!-\!\tfrac{1}{n},l_{m}\!-\! \tfrac{1}{m}\)\right]\nonumber\\
    &\times \left[C\(k_{n},l_{m}\)+C\(k_{n},l_{m} \!-\! \tfrac{1}{m}\) + C\(k_{n} \!-\!
        \tfrac{1}{n},l_{m}\) +
        C\(k_{n}\!-\!\tfrac{1}{n},l_{m}\!-\!\tfrac{1}{m}\)\right],\nonumber
    \\A_2 &=\lim_{n,m\rightarrow\infty}\frac{1}{4} \left[C\(k_{n},k_{n}\) +
        C\(k_{n},k_{n}\!-\!\tfrac{1}{n}\) - C\(k_{n}\!-\!\tfrac{1}{n},k_{n}\) -
        C\(k_{n}\!-\!\tfrac{1}{n},k_{n}\!-\!\tfrac{1}{n}\)\right] \nonumber\\
    &\times \left[C\(l_{m},l_{m}\)+C\(l_{m},l_{m} \!-\!
        \tfrac{1}{m}\)-C\(l_{m}\!-\!\tfrac{1}{m},l_{m}\) -
        C\(l_{m}\!-\!\tfrac{1}{n},l_{n}\!-\!\tfrac{1}{n}\)\right],
    \nonumber\\
    A_3 &=\lim_{n,m\rightarrow\infty}\frac{1}{4}\left[C\(k_{n},l_{m}\)+C\(k_{n},l_{m}
        \!-\! \tfrac{1}{m}\)C\(k_{n}\!-\!\tfrac{1}{n},l_{m}\) -
        C\(k_{n}\!-\!\tfrac{1}{n},l_{m}\!-\!\tfrac{1}{m}\)\right] \nonumber\\
    &\times \left[C\(l_{m},k_{n}\)+C\(l_{m},k_{n} \!-\! \tfrac{1}{n}\)-C\(l_{m} \!-\!
        \tfrac{1}{m},k_{n}\)-C\(l_{m}\!-\!\tfrac{1}{m},k_{n}\!-\!\tfrac{1}{n}\)\right].
\la{A.2}
\end{align}
This results from the application of Wick's theorem to the Gaussian average
(\ref{A.1}) with covariance (\ref{3.2}). Expanding
$C\(k_{n}-\tfrac{1}{n},l_{m}\)=C\(k_{n},l_{m}\)-\tfrac{1}{n}(\partial_{1}C)\(k_{n},l_{m}\)$
and $C\(k_{n},l_{m}-\tfrac{1}{m}\)=C\(k_{n},l_{m}\)-\tfrac{1}{m}
(\partial_{2}C)\(k_{n},l_{m}\)$ and taking the limits $n,m\to\infty$ gives
\begin{align}
    A_1 &=\int_{0}^{1}\!\!\! \d s \!\int_{0}^{1}\!\!\!\d t
    \ C(s,t)(\partial_{1}\partial_{2}C)(s,t) =\frac{1}{12},\nonumber\\
    A_2 &=\frac{1}{4}\(\int_{0}^{1}\!\!\! \d s \,\frac{\d}{\d s}\,
    C(s,s)\)^{2}=0,\nonumber\\
    A_3 &=\int_{0}^{1}\!\!\!\d s \!\int_{0}^{1}\!\!\! \d
    t\ (\partial_{1}C)(s,t)\, (\partial_{2}C)(s,t) =-\frac{1}{12},
\la{A.3}
\end{align}
hence the result (\ref{B.31}).

\section*{Appendix B}

From (\ref{B.14}) and (\ref{B.19}) the cross Coulomb-magnetic term is
\begin{align}
    & \beta^{2}e_a^2 e_b^2\int\!\! \D(\b\xi_{a})\rho(\b\xi_{a})\!
    \int\!\! \D(\b\xi_{b})\rho(\b\xi_{b})\ \Wc(\mathcal{F}_{a},\mathcal{F}_{b})
    \Wm(\mathcal{F}_{a},\mathcal{F}_{b})
    =\nonumber\\
    &\frac{\beta e_a^2 e_b^2}{\sqrt{ m_{a}m_{b}}c^{2}} \int\!\!\!
    \frac{\d\bk_{1}}{(2\pi)^{3}}\! \int\!\!\! \frac{\d\bk_{2}}{(2\pi)^{3}}\,
    \e^{i(\bk_{1}+\bk_{2})\cdot(\r_{a}-\r_{b})} \!\int_{0}^{1}\!\!\! \d s_{1}
    \!\int_{0}^{1}\!\!\! \d s_{2}\,\big(\delta(s_{1}-s_{2})-1\big)
    \nonumber\\
    & \quad\quad\times \big(H_{a}^{\mu}\big)^\ast(\bk_{1},\bk_{2},
    s_{1})H_{b}^{\nu}(\bk_{1},\bk_{2},s_{2})\, \frac{4\pi}{k_{1}^{2}}\,
    G^{\mu\nu}(\bk_{2}),
\la{D.1}
\end{align}
where
\begin{align}
    H_{a}^{\mu}(\bk_{1},\bk_{2}, s_{1})=\int\!\! \D(\b\xi_{a})\rho(\b\xi_{a})
    \,\e^{-i \lambda_{a}\bk_{1}\cdot\b\xi_{a}(s_{1})} \!\int_{0}^{1}\!\!\!
    \d\xi_a^{\mu}(s)\ \e^{-i \lambda_{a}\bk_{2}\cdot\b\xi_{a}(s)}.
\end{align}
Because of the rotational invariance of $\D(\b\xi_{a})\rho(\b\xi_{a})$, averages
of odd powers of $\b\xi_{a}$ vanish. This implies that in the small-$\bk_{1},
\bk_{2}$ expansion of $ H_{a}^{\mu}(\bk_{1},\bk_{2}, s_{1})$ only odd monomials
in $\bk_{1}, \bk_{2}$ occur:
\begin{align}
    H_{a}^{\mu}(\bk_{1},\bk_{2}, s_{1}) &\sim \int\!\!
    \D(\b\xi_{a})\rho(\b\xi_{a}) \!\int_{0}^{1}\!\!\!
    \d\xi_a^{\mu}(s)\ \big(i\lambda_{a}\bk_{2}\cdot\b\xi_{a}(s)\big) +
    \Order_3(\bk_{1},\bk_{2}) \nonumber
    \\ &= \text{const}\times k_{2}^{\mu} + \Order_{3}(\bk_{1},\bk_{2}),
\label{D.2}
\end{align}
where $\Order_{3}(\bk_{1},\bk_{2}) $ represent monomials of order 3 in the
components of $\bk_{1},\bk_{2}$. The same holds for $
H_{b}^{\mu}(\bk_{1},\bk_{2}, s_{2})$. Since $k_{2}^{\mu}G^{\mu\nu}(\bk_{2})=0$
by transversality, one concludes that the term (\ref{D.1}) decays at least as
$|\r_{a}-\r_{b}|^{-8}$.

\section*{Appendix C}

If point $i$ in $\Wc(i,j)$ or $\Wm(i,j)$ is a classical end, there is no
other $\bxi_i$ dependence at this point than that arising from these
bonds. In the asymptotic formula (\ref{B.21}) for $\Wc$, this dependence is
linear and thereby vanishes upon the space-inversion invariant
$\D(\bxi_i)$ integration. In the case of $\Wm$, from formula (\ref{B.14}),
the $\D(\bxi_i)$ integration yields the factor
\begin{align}
	\int \!\!\D(\bxi_i) \!\int_0^1\!\!\! \d\xi^\mu_i(s_1)\ \e^{i \k \cdot
    \lambda_i\bxi_i(s_1)} \propto k^\mu
\end{align}
because of covariance under rotation. Hence, this contribution vanishes as a
consequence of transversality $k^\mu G^{\mu\nu}(\k)=0$.

\section*{Appendix D}
An explicit expression for the constant $a=a(\hbar, \beta, \rho)$ follows
from taking the trace in equation (\ref{F.15}) and using (\ref{F.13})
expanded for small $k$. This yields
\begin{align}
	a=&\frac{1}{2}\sum_{\gamma, \gamma'} \frac{e_\gamma \lambda_\gamma
        e_{\gamma'} \lambda_{\gamma'}}{\beta \sqrt{m_\gamma m_{\gamma'}}
        c^2}\int\!\!\D(\bxi)\!\!\! \int\!\!\D(\bxi') \nonumber
	\\&\times\int_0^1 \!\!\! \d\xi^\mu(s) \!\!\! \int_0^1 \!\!
    \d\xi^\nu(s') \big(\hat \k \cdot \bxi(s)\big) \big(\hat \k \cdot
    \bxi'(s')\big) \delta_\text{tr}^{\mu\nu}(\hat \k) \,
    n_\text{T}(\gamma,\bxi,\gamma',\bxi',\k=\mathbf{0}),
\end{align}
where $\hat \k=\k/k$. As $\lambda_\gamma \lambda_{\gamma'}$ is of order
$\hbar^2$, at lowest order in $\hbar$ one can set $\hbar=0$ in the
correlation function. The latter becomes independent of the quantum
fluctuations $\bxi, \bxi'$ and reduces to the density correlation function
of the corresponding classical system. The remaining functional integrals,
of the type $\int\D(\bxi) \int_0^1 \d \xi^\mu(s) \xi^\sigma(s)$, vanish
identically. The terms of order $\Order(\hbar)$ in $n_\text{T}$ are
necessarily linear in $\bxi$ or $\bxi'$. They do not contribute to $a$
since averages of odd powers $\bxi$ or $\bxi'$ are zero, implying that
there are no $\hbar^3$-terms in $a$. We thus conclude that $a$ is
$\Order(\hbar^4)$.

\end{document}